\documentclass[preprint2]{aastex}

\usepackage{aastex_plus}
\shorttitle{Magnetic Field and Chromospheric Emission in an Active Region}
\shortauthors{Brooks et al.}
\begin{document}

\title{Characteristics and Evolution of the Magnetic field and Chromospheric Emission in an Active Region Core Observed by {\it Hinode}}
\author{David H. Brooks \altaffilmark{1,2,3}, Harry P. Warren \altaffilmark{1}, Amy R. Winebarger \altaffilmark{4}}
\altaffiltext{1}{Space Science Division,  Code 7673, Naval Research Laboratory, Washington, DC 20375}                       
\altaffiltext{2}{George Mason University, 4400 University Drive, Fairfax, VA 22020}                                  
\altaffiltext{3}{Present address: Hinode Team, ISAS/JAXA, 3-1-1 Yoshinodai, Sagamihara, Kanagawa 229-8510, Japan}
\altaffiltext{4}{Department of Physics, Alabama A\&M, 4900 Meridian Street, Normal, AL 35762} 
\email{dhbrooks@ssd5.nrl.navy.mil}

\begin{abstract}
We describe the characteristics and evolution of the magnetic field and chromospheric 
emission in an active
region core observed by the Solar Optical Telescope (SOT) on {\it Hinode}. Consistent with
previous studies, we find that the
moss is unipolar, the spatial distribution of magnetic flux evolves slowly, and
that the magnetic field is only moderately inclined. 
We also show that the 
field line inclination and
horizontal component are coherent, and that the magnetic field is
mostly sheared in the inter-moss regions where the highest magnetic flux
variability is seen. 
Using extrapolations from Spectropolarimeter (SP) magnetograms we show that
the magnetic connectivity in the moss is different than in the quiet Sun
because most of the magnetic field extends to significant coronal heights.
The magnetic flux, field vector, and chromospheric emission in the moss also appear highly
dynamic, but actually show only
small scale variations in {\it magnitude} on time-scales 
longer than the cooling times for hydrodynamic loops computed from our 
extrapolations, suggesting high-frequency (continuous) heating events. 
Some evidence is found for flux (\ion{Ca}{2} intensity) changes on the order of 100--200 G (DN) on time-scales of
20--30 mins that could be taken as indicative of low-frequency heating. 
We find,
however, that only a small fraction (10\%) of our simulated loops would be expected to 
cool on these time-scales, and we do not find clear
evidence that the flux changes consistently produce intensity changes in the chromosphere.
Using observations
from the EUV Imaging Spectrometer (EIS) we
also determine that the filling factor in the moss is $\sim$ 16\%, consistent
with previous studies and larger than the size of an SOT pixel. The magnetic
flux and chromospheric intensity in most individual SOT pixels in the moss vary by less than $\sim$ 20\% and $\sim$10\%, respectively, on loop cooling
time-scales. 
In view of the high energy requirements of the chromosphere, we suggest that these variations could 
be sufficient for the heating of `warm' EUV loops, but that the high basal levels may be more
important for powering the hot core
loops rooted in the moss.
The magnetic field and chromospheric emission appear to evolve gradually
on spatial scales comparable to the cross-field scale of the fundamental coronal structures
inferred from EIS measurements. 

\end{abstract}
\keywords{Sun: activity--- Sun: magnetic topology---Sun: photosphere---Sun: chromosphere---Sun: corona}             

\bibliographystyle{/data2/latex/dhb_bib/apj}

\section{Introduction}
Significant progress in solving the decades old coronal heating problem could be
made if one knew the duration
and frequency of heating events. Analysis of soft X-ray loops in the {\it Yohkoh}
era suggested that high temperature (3--5MK) coronal plasma could be heated steadily
\citep{porter&klimchuk_1995,kano&tsuneta_1996}. These loops are rooted in the ``moss,'' which is found in
active region cores \citep{martens_etal2000}, and it has been argued that the lack of intensity variations
there is indicative of steady heating \citep{antiochos_etal2003}. 
Furthermore, hydrostatic modeling of whole active regions has been quite successful
at reproducing the core emission from short hot loops \citep{schrijver_etal2004,warren&winebarger_2006,lundquist_etal2008}.

Hydrostatic modeling, however, has greater difficulty reproducing the emission at
lower temperatures. Warm (1MK) EUV loops observed by the Solar and Heliospheric Observatory
\citep[{\it SOHO},][]{domingo_etal1995} and the Transition Region and Coronal Explorer \citep[{\it TRACE},][]{handy_etal1999} have been found to be 
overdense compared to static equilibrium
theory and persist far longer than expected loop cooling times \citep{lenz_etal1999,aschwanden_etal2001,winebarger_etal2003a}. 
These observational features can be explained if coronal loops are bundles of
unresolved threads that are heated impulsively \citep{aschwanden_etal2000b,warren_etal2002,winebarger_etal2003b}. 

Of course, most of the proposed coronal heating mechanisms are impulsive in nature \citep{klimchuk_2006},
and the term `steady heating' is usually taken to mean that the repetition time between
impulsive heating events is shorter than the time it takes for the loop to cool by conduction and then radiation. 
Loops are thus maintained at high temperatures and the
emission is apparently steady. 
Many active regions appear to evolve slowly and loops
are not often seen cooling in the core around the moss in these regions \citep{antiochos_etal2003,patsourakos&klimchuk_2008}. In other
cases, loops are clearly seen evolving and cooling \citep{ugarteurra_etal2009},
and it is not clear which type of heating is dominant.
Still, it would be surprising if the structure and heating characteristics of warm loops and hot loops were fundamentally
different. 
A challenge to current loop modeling is to draw these pictures together and 
understand how they can exhibit these apparently contradictory properties. 

The instruments on board the {\it Hinode} satellite \citep{kosugi_etal2007} are providing unprecedented observations
of active regions in terms of temperature coverage, spatial, temporal, and spectral 
resolution. As such they are allowing us to probe the properties of moss at the bases
of high temperature loops in active region cores
in new detail. In two previous papers 
\citep[hereafter Paper 1 and Paper 2]{brooks&warren_2009,warren_etal2010a}, we analyzed 
EUV Imaging Spectrometer (EIS)
and X-ray Telescope (XRT) observations of an active region observed by {\it Hinode} in June 2007
(NOAA 10960) in order to study the time-scale of energy release. Several lines of evidence led to the
conclusion that the heating in the core of this region was effectively steady. First, soft X-ray 
and \ion{Fe}{12} 195.119\,\AA\, intensities in the moss were shown to vary by less than 15\% over many
hours. Second, from the \ion{Fe}{12} 195.119\,\AA\, line profiles, we measured Doppler and non-thermal
velocities in the moss of $\sim$ 3 km s$^{-1}$ and $\sim$ 26 km s$^{-1}$ on average, respectively. Taking
into consideration the uncertainties, and also the fact that \ion{Fe}{12} lines have been shown to be blue-shifted by a 
few km s$^{-1}$ in the quiet Sun \citep{peter&judge_1999}, the moss velocity measurement is consistent
with zero. In addition, based on a comparison with several quiet Sun synoptic datasets, the measurement 
of the non-thermal velocity in the moss was shown to be
no larger than the typical quiet Sun value of 25 km s$^{-1}$. More importantly, neither quantity varied
by more than 15\% over many hours. Third, no evidence was found for co-spatial warm and hot
emission, as would be expected from impulsive heating models that assume coronal loops have time
to cool substantially between events. Finally, the observed moss intensities could be brought into agreement
with hydrostatic simulations provided that local expansion of the magnetic field at the base of the
corona was included in the numerical model.

These results 
rule out the possibility of low frequency impulsive heating of monolithic loops on
the spatial scales resolved in the analysis. An alternative possibility, however, 
is that the heating
is impulsive at low frequency on sub-resolution threads,
as is thought to be the case in the warm overdense loops (Jim Klimchuk 2009, private communication).
This picture may appear like high frequency heating in the observational analysis of Paper 1.

It is clear that we can already
resolve the expected loop cooling times temporally, and even other transient activity on localized 
scales, e.g. short duration ``Type II'' spicules \citep{depontieu_etal2007} and blinkers \citep{brooks_etal2004}. Some
recent observations by EIS have also suggested that we may be close to resolving the cross field
spatial scale of the fundamental coronal structures: see e.g. \citealt{warren_etal2008a} who derive loop filling
factors of $\sim$ 10\%, or \citet{tripathi_etal2009} who obtained values larger than this except near the base
of the loop they analyzed. This also sets an upper limit on the spatial scale of coronal heating in loops,
whether the fundamental structures are monolithic or composed of multiple threads. 
It does not mean, however, that the individual threads of the bundle are resolved.
Based on magnetic flux arguments, \citet{priest_etal2002} suggest that single {\it TRACE} loops
may be composed of 10 or more individual strands, and that the spatial scale we need to observe may 
be finer still. This is the case for nanoflare heating \citep{parker_1988}
where the reconnection takes place at the current sheets between threads \citep{klimchuk_2006}.
Current EUV and X-ray instrumentation are unable to
resolve such structure, but the spatial resolution, excellent seeing, and stability of the
{\it Hinode} Solar Optical Telescope (SOT) allow it to observe at much higher spatial resolution than 
EIS or XRT. A direct comparison between chromospheric and coronal observations is of course difficult
because of the large difference in instrumental spatial resolutions, and a lack of knowledge of the
influence of the expansion of the field.
Therefore, in this paper, we analyze only SOT observations of the magnetic field and chromospheric emission.
We again study the core of the June 2007 region analyzed in Paper 1 and Paper 2. Since the observations
are new in themselves, we describe the magnetic field characteristics of the moss and active
region core. Our main goal, however, is to study the variation of the magnetic flux, vector field,
and chromospheric emission
to try to uncover any behavior that could be related to the heating process and would allow us
to set further constraints on the heating time-scale in this region.

In \S \ref{obs} we describe the observations
and data reduction procedures. In \S \ref{characterB} we discuss the magnetic characteristics of
the region and in \S \ref{model} we introduce the modeling that we use to compare the time-scales
of observational variability with typical loop cooling times. This is important because it is the
crucial time-scale that differentiates between low and high frequency (effectively steady) impulsive 
heating. We thus put several statements about
the evolution of the magnetic field on a quantitative basis. In \S \ref{variability} we examine
the variability and evolution of the magnetic field and the chromospheric emission, and find that most of the activity
takes place on time-scales longer than the computed theoretical loop cooling times. We also discuss
departures from this picture.
In \S \ref{ff} we discuss the spatial resolution issue once again and we 
derive the filling factor for the moss using EIS observations. The results suggest that SOT
{\it may} be able to resolve the cross-field spatial scale of structures in the moss in individual pixels, so we further 
examine the magnetic flux and chromospheric variability on these size scales \S \ref{individual}. 
The conclusions are presented and discussed in 
\S \ref{discuss}.

\section{{\it Hinode} and {\it TRACE} Observations}
\label{obs}
AR 10960 crossed the solar disk between 2007, May 30 and June 14. The region
produced numerous C- and M- class flares during that period and was therefore
the main observing target for most solar instrumentation on the ground and
in space. In this paper we mostly use SOT data, but we also coalign and use
{\it TRACE} 171\,\AA\, filter images to identify the coronal features in the active
region core, particularly the moss. An XRT Open/Ti-Poly image is also used for
giving an overview of the hot core emission, and later we also use EIS data to determine the 
moss filling factor.

The SOT is described in detail by \citet{tsuneta_etal2008}. It consists of 
an optical telescope assembly \citep[OTA,][]{suematsu_etal2008} that feeds 
a Filtergraph (FG) instrument and Spectropolarimeter (SP). The FG itself 
consists of a
Broadband Filter Imager (BFI) and a tunable Lyot-type Narrowband Filter Imager (NFI).
The NFI can obtain filtergrams, dopplergrams, and Stokes I, Q, U, and V images in a
number of spectral lines formed in the solar photosphere and chromosphere. 
The SP instrument obtains
high-precision polarimetric scans in the Fe {\sc i} 6301\,\AA\, and 6302\,\AA\, spectral lines.
The precision of the
polarimetric calibration is discussed in 
\citet{ichimoto_etal2008}. As it is flown in space and has its own correlation tracker, SOT
obtains high quality seeing-free and stable longitudinal or transverse magnetograms that 
can be used to diagnose
magnetic field dynamics in the lower atmosphere. 

%%%%
\begin{figure*}
\centering
\includegraphics[width=0.45\linewidth]{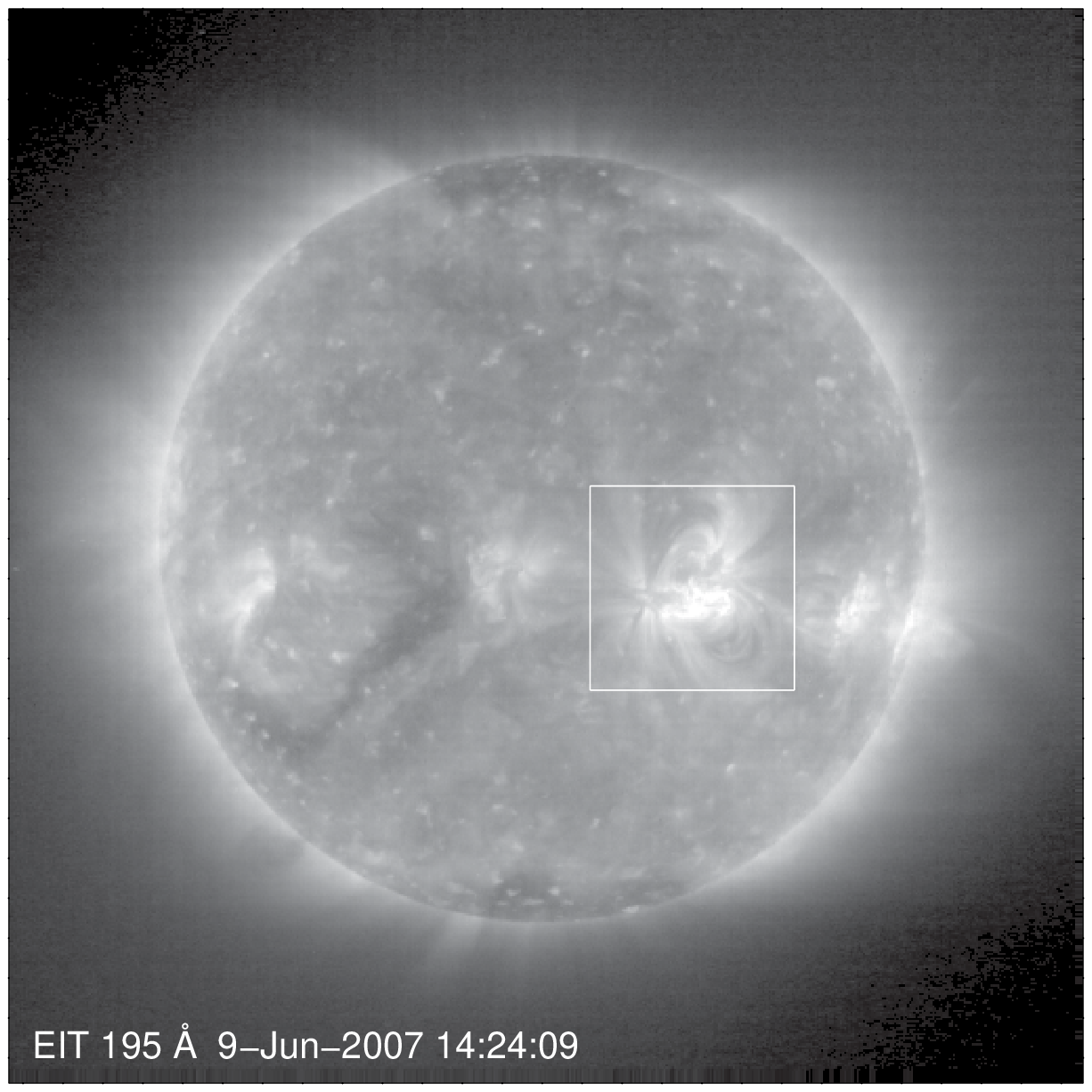}
\includegraphics[width=0.45\linewidth]{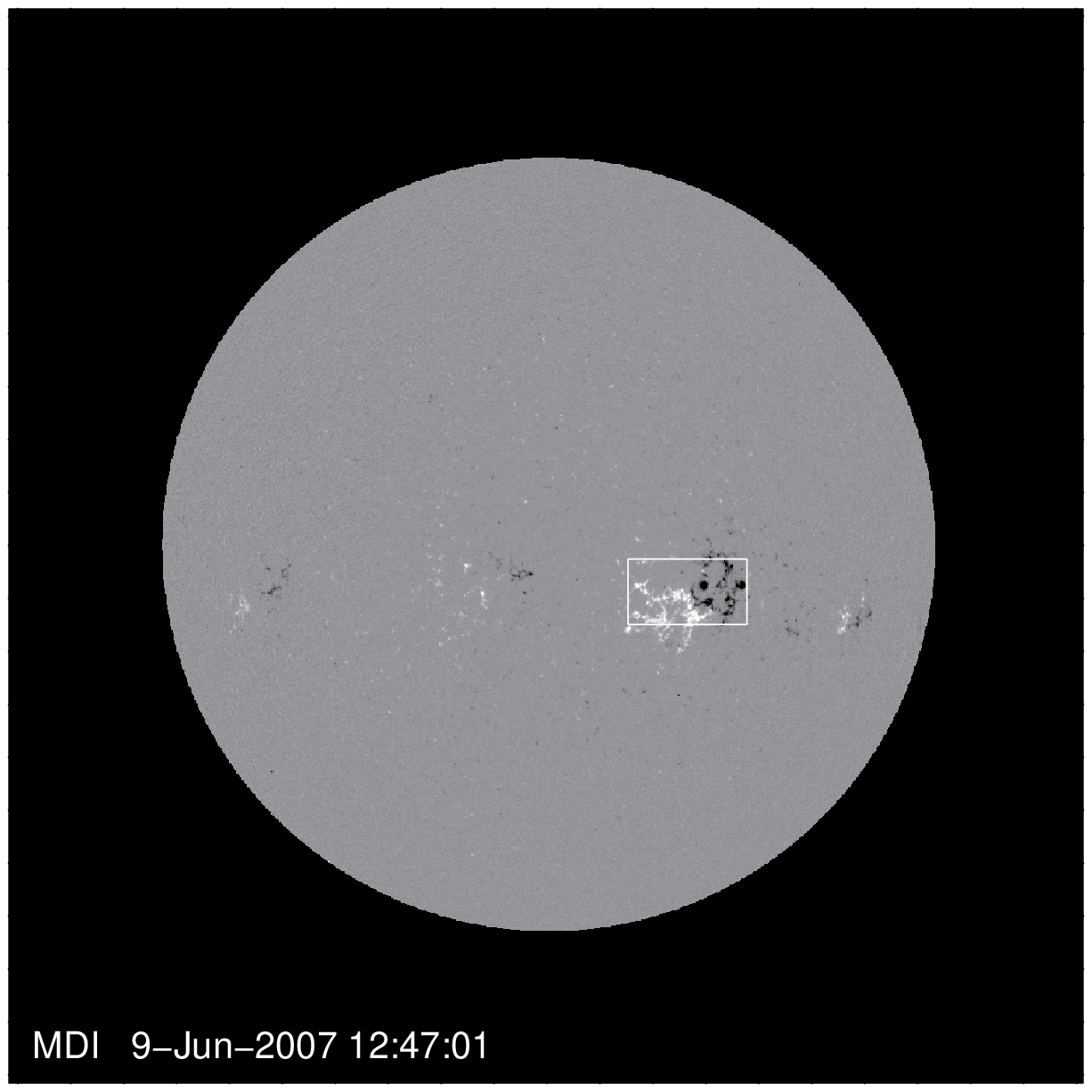}
\caption{Full Sun context images of AR 10960. Left Panel: {\it SOHO}/EIT 195\,\AA\, image with the {\it TRACE} FOV overlaid as a box. Right Panel: {\it SOHO}/MDI magnetogram with the {\it Hinode} SOT FOV overlaid as a box. 
\label{fig1}}
\end{figure*}
In this paper we analyze both FG and SP data. Large FOV SP observations are used in
\S \ref{characterB} to describe the magnetic characteristics of AR 10960. These data
were obtained from the SOT level-2 archive. As such they are outputs from the 
Milne-Eddington gRid Linear Inversion Network (MERLIN) code developed at the 
Community Spectro-polarimetric Analysis Center (CSAC) at the High Altitude 
Observatory (HAO) by Bruce Lites and colleagues \citep{lites_etal2007}. MERLIN
performs Levenberg-Marquardt least squares fitting of the full Stokes profiles
obtained by the SP. Several assumptions about the atmospheric approximation and
fit parameter initialization are made, for example, the source function is assumed
to be linear with optical depth and the atmosphere is assumed to be in local
thermal equilibrium. The level-2 archive data are the parameters that give the 
best fit to the observed profile and much more detail is given on the level-2
archive website. A key point of note is that no attempt is made to resolve the 
180$^o$ azimuth ambiguity. This is discussed at the appropriate times below. 

\begin{figure}[ht]
\centering
\includegraphics[width=0.88\linewidth,viewport= 50 40 432 216]{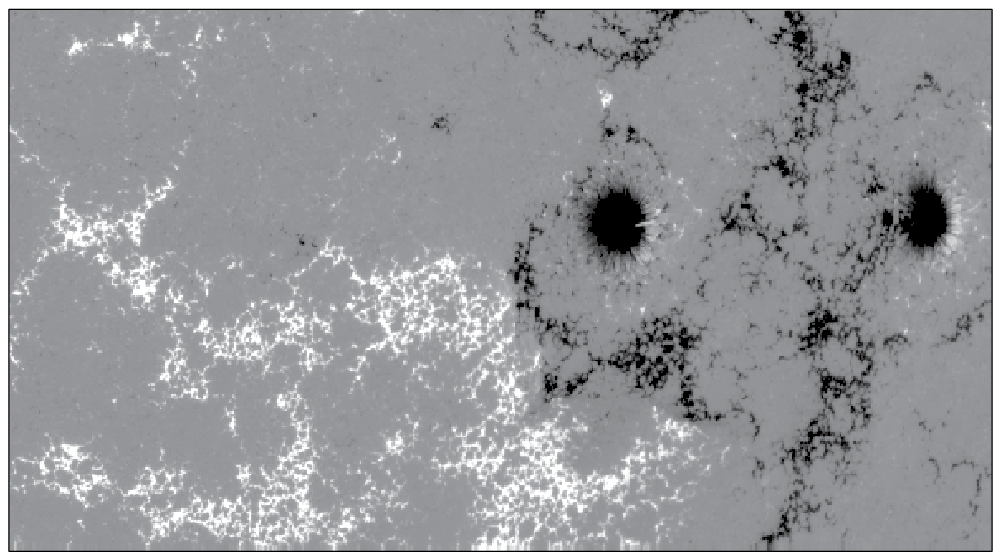}
\includegraphics[width=0.88\linewidth,viewport= 50 40 432 216]{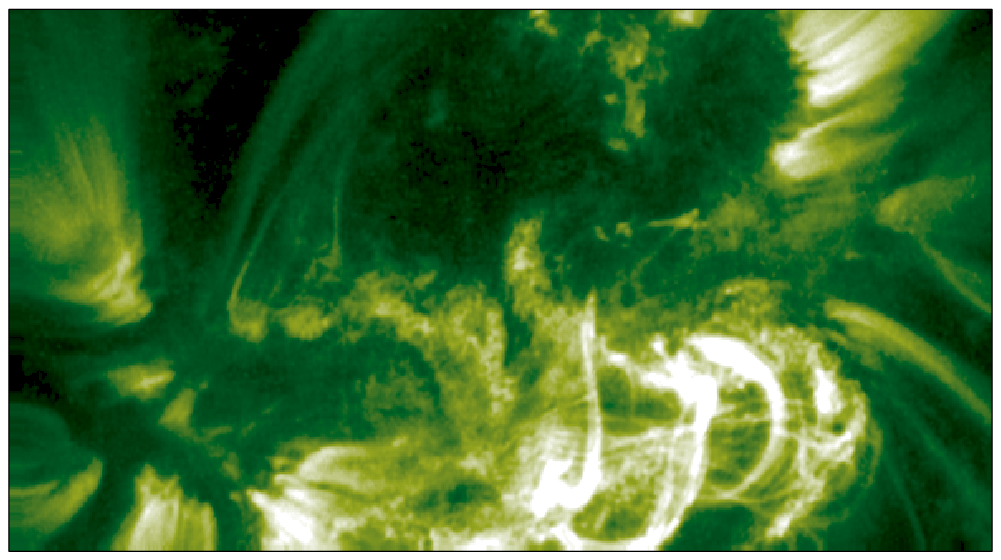}
\includegraphics[width=0.88\linewidth,viewport= 50 40 432 216]{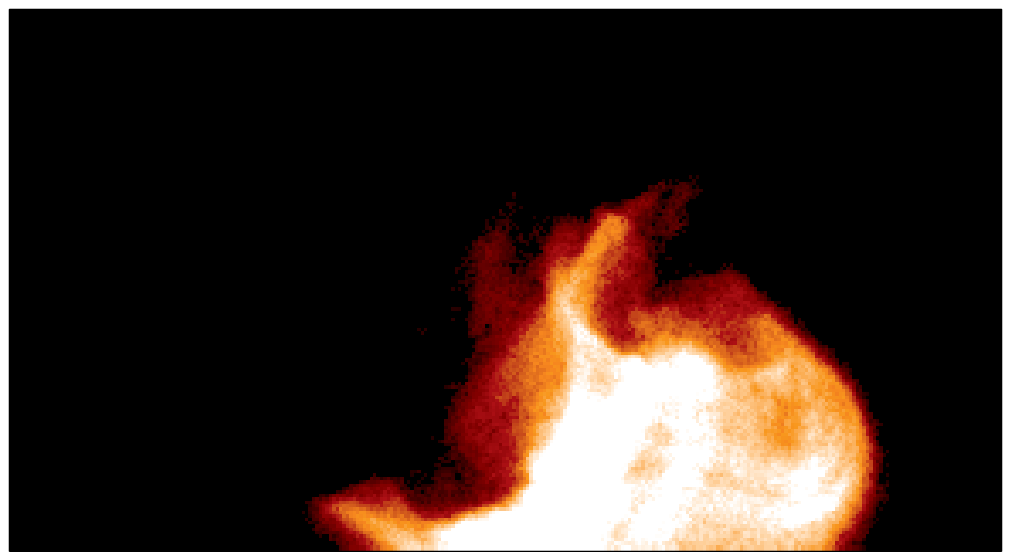}
\caption{SP, {\it TRACE}, and XRT images of AR 10960. Top Panel: {\it Hinode} SOT/SP map corresponding to the FOV shown in Figure \ref{fig1}. The image shows the magnitude
of Stokes V signal in the \ion{Fe}{1} 6302\,\AA\, profile. Middle Panel: Coaligned {\it TRACE} 171\,\AA\, filter image taken at 14:37UT during the SP scan and showing the moss emission. 
Approximately coaligned XRT image taken at 14:35UT during the SP scan and showing the location
of the high temperature core emission. 
\label{fig2}}
\end{figure}
The large FOV SP observations we analyze were obtained by scanning the slit over
an area of 279.1$'' \times$163.8$''$ between 14:20:05 and 15:23:17UT on June 09. 
The resolution of this scan is 0.30$'' \times$0.32$''$ per pixel and the 
time for individual polarimetric exposures was 1.6s. Exposures were obtained at
each new scan position every 3.8s.
In \S \ref{variability} we analyze a time-series of high cadence SP scans obtained
between 21:56:03 and 22:59:31UT on June 08. The observing parameters were the same
for this run, but the FOV was only 7$'' \times$512$''$ so the time for each small scan
is 28s. These data were also obtained from the level-2 archive.

We also analyze a time-series of FG longitudinal magnetograms obtained in 
the \ion{Na}{1} D 5896\,\AA\, line at $\sim$ 30s cadence. The spatial sampling of
the \ion{Na}{1} D Stokes images was 0.16$''$/pixels over a 327.7$'' \times$163.8$''$ FOV
with an effective exposure time of 14.2s created from polarimetric images with
individual exposure times of 0.12s.
The time-series ran between 18:14:32 and 23:37:04UT on June 09, and the data were
processed using the {\it SolarSoft} routine FG\_PREP. The SOT observations were 
set to complement each other, with similar FOV scans with the SP and FG interspersed
with high cadence time-series. They cover approximately 26 hr in total.

To study the variability of chromospheric emission we also analyze a co-temporal FG
time-series of \ion{Ca}{2} 3896\,\AA\, images taken between 18:14:54 and 23:37:26UT
on June 09. Images were obtained at 30s cadence with an exposure time of 0.15s. 
The spatial sampling of these data is 0.109$''$/pixels over a 111.6$'' \times$111.6$''$
FOV. To estimate the statistical noise in \ion{Ca}{2} in \S \ref{ca_var} we use
a very high cadence (8s) time-series of a plage region obtained on 2007, February 20,
between 11:57 and 13:37UT. All the \ion{Ca}{2} data were processed using FG\_PREP.

{\it TRACE} also repeatedly observed AR 10960 during its passage across the disk, and 
movies of the region have been presented in Paper 1 and Paper 2. In \S \ref{characterB} 
we use a 171\,\AA\, filter image obtained at 14:37:23UT on June 09 to identify the moss
areas. The FOV was 512$'' \times$512$''$ and the exposure time was 32.8s. We also use
a 1600\,\AA\, filter image obtained at 14:36:09UT to coalign with the SP data. The
exposure time for this image was 0.86s. The data 
were processed and despiked using the {\it SolarSoft}
routine TRACE\_PREP. In addition, we use an XRT Open/Ti-Poly image taken at 14:34:46UT
with a 512$'' \times$512$''$ FOV and 0.06s exposure time. This image was processed
using XRT\_PREP.

In section \S \ref{ff} we use EIS data to measure the moss filling factor. The
EIS instrument is described in detail by \citet{culhane_etal2007a} and \citet{korendyke_etal2006}. 
It observes
with high spectral and spatial resolution (22.3m\AA\, and 1$''$ pixels, respectively)
in short and long wavelength bands in the ranges 171--212\,\AA\, and 245-291\,\AA.
Several slits from 1$''$ to 266$''$ wide are available for making observations. 
In this paper we use a raster scan obtained with the observing sequence AR\_velocity\_map
at 10:58:10UT on June 09. The sequence runs for about 5 hr 15 mins and takes a context 
image by stepping the 40$''$ slit 
across a large FOV followed by a 1$''$ slit scan over 330$'' \times$304$''$.
We only use the 1$''$ slit scan data here, the exposure time for which was 40s at 
each position. Many lines are included in the study, though we only need one 
density diagnostic line pair for our analysis. 
The data were processed using the default options in the 
{\it SolarSoft} routine EIS\_PREP. Each of the lines used were
then fitted with Gaussian profiles to obtain the line intensity. Further details
are given in \S \ref{ff}.

\section{Magnetic Characteristics of AR 10960}
\label{characterB}
Here we describe the overall magnetic structure of AR 10960.
To understand the relationship between the moss and the magnetic field we need
to coalign the {\it TRACE} image to the SP large FOV map. To do this, 
we first cut out the common area from the {\it TRACE} 1600\,\AA\, image
as determined from the FITS header coordinates. We then re-sampled the SP data
to the lower {\it TRACE} resolution (note that all of the quantitative analysis is
done on the original data), 
and then coaligned the 1600\,\AA\, image with the SP
measured non-directional magnetic field strength. 
From this procedure we corrected the uncertainty in the common
area of the 1600\,\AA\, image determined from the FITS header coordinates. 
We then extracted the correct area from the 1600\,\AA\, image. 

This procedure worked well, but some stretching and rotation of the 1600\,\AA\,
image compared to the SP map was evident as a result of the difference
in plate scale magnification and satellite orbital attitude. These effects were corrected
after visual inspection. 
The rotation correction is approximately 1$^o$ counterclockwise
and the magnification correction is approximately 2\% between the {\it TRACE} and
SOT images. Furthermore, an E-W pixel shift of about 2 pixels was identified and removed.

\begin{figure}[ht]
\centering
\includegraphics[width=0.88\linewidth,viewport= 50 40 432 216]{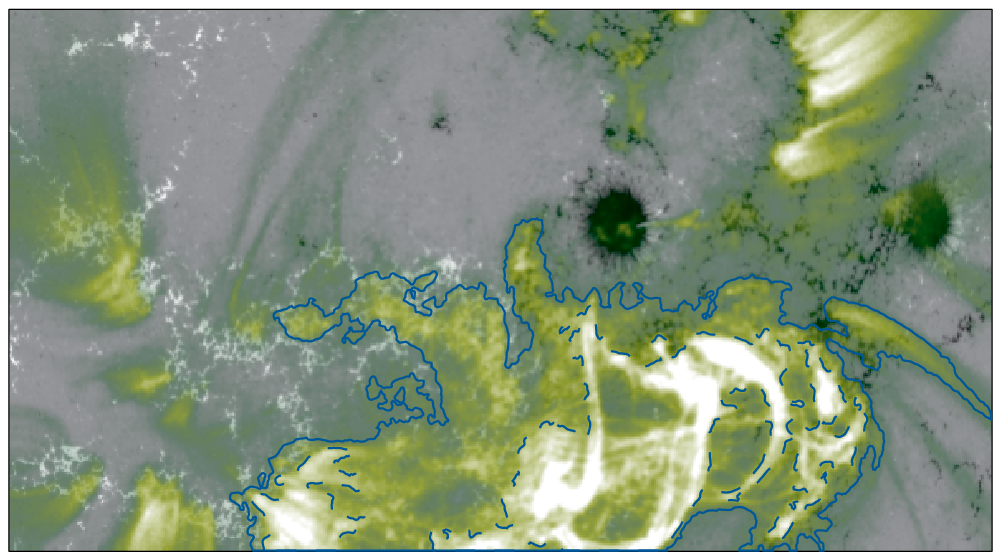}
\includegraphics[width=0.88\linewidth,viewport= 50 40 432 216]{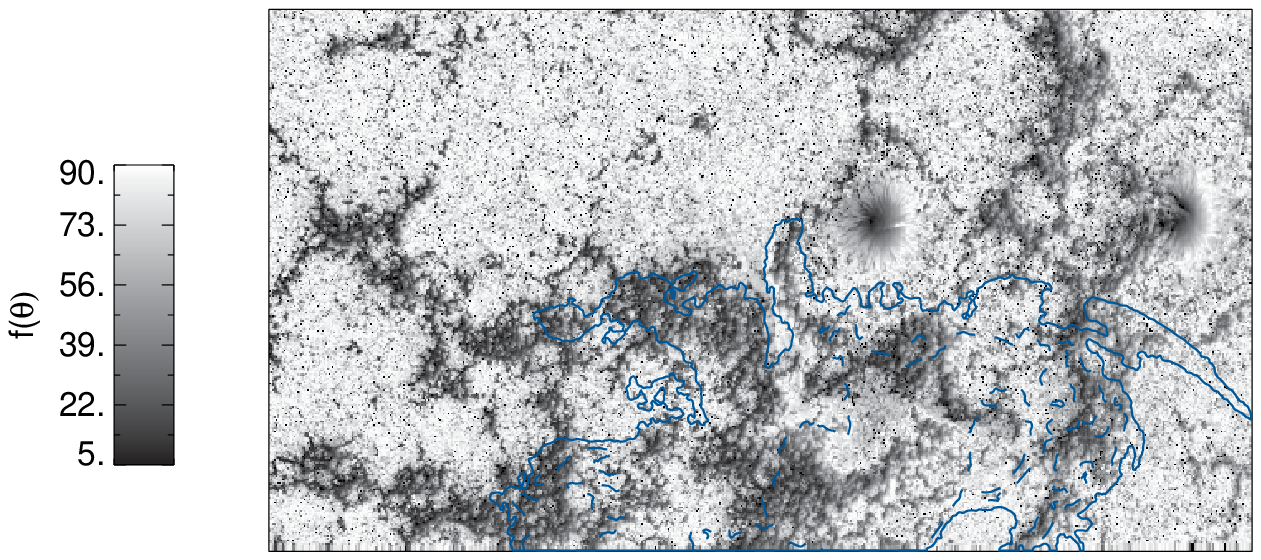}
\includegraphics[width=0.88\linewidth,viewport= 50 40 432 216]{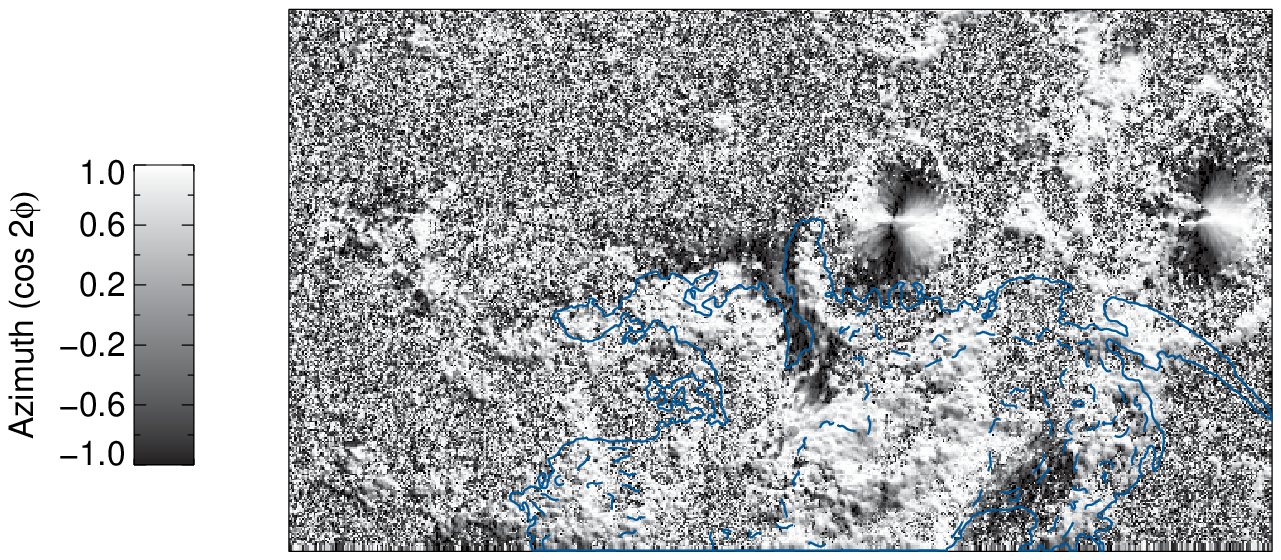}
\caption{Coaligned SP and {\it TRACE} overlay, magnetic field inclination, and azimuth angle 
Top Panel: {\it TRACE} image overlaid in green on the SOT/SP map. The {\it TRACE} image has been scaled logarithmically. The solid blue line outlines the major moss region and the dashed blue line is intended to exclude obvious loop emission.
Middle Panel: $f (\theta)$, where $\theta$ is the inclination angle. Lower Panel: $\cos 2\phi$, where $\phi$ is the azimuth angle. The same contours from the upper panel are overlaid in blue on the middle and lower panels.
\label{fig3}}
\end{figure}
Having established the alignment between the SOT and 1600\,\AA\, images, the coalignment
with the 171\,\AA\, {\it TRACE} image was finally made by cross-correlating the two 
{\it TRACE} images after the 171\,\AA\, image had been corrected for the inter-{\it TRACE}
filter offsets reported by \citet{handy_etal1999}. 

Figure \ref{fig1} 
shows full disk images taken on June 09 by the {\it SOHO} Extreme
ultraviolet Imaging Telescope \citep[EIT,][]{delaboudiniere_etal1995} and
Michelson Doppler Imager \citep[MDI,][]{scherrer_etal1995}. The FOV of the
{\it TRACE} data used to coalign with SOT is shown on the EIT image, and the
FOV of the SP large FOV scan is shown on the MDI image. The extent of the
active region and magnetic configuration are easily seen. Note the relative
lack of complex emission or obscuration of the moss in the core of the region.
This makes this particular region an ideal candidate for study.

\begin{figure}[ht]
\centering
\includegraphics[width=0.98\linewidth,viewport= 60 60 360 360]{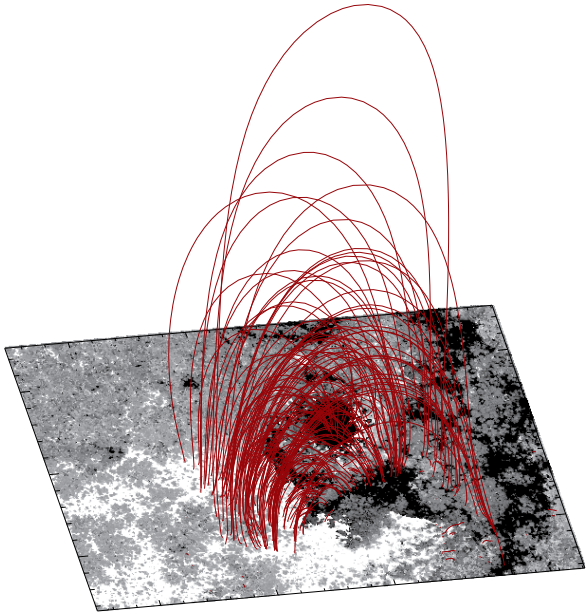}
\caption{Potential field extrapolation from the moss region indicated in
Figure \ref{fig2}. Only 218 field lines are shown for clarity.
\label{fig4}}
\end{figure}
Figure \ref{fig2} shows the magnitude of Stokes V signal measured in the 
\ion{Fe}{1} 6302\,\AA\, line, a coaligned 171\,\AA\, {\it TRACE}
image, and an XRT Open/Ti-Poly image. 
The region is $\beta\gamma$ by June 09, with well separated positive and 
negative polarity flux. There
is strong moss emission above both polarities in the core, separated by a dark channel
above weak magnetic flux. As pointed out previously by others 
\citep{katsukawa&tsuneta_2005,tripathi_etal2008,brooks_etal2008,title_etal2009},
the plage and moss areas are predominantly unipolar with unmixed flux, though the spatial correlation 
between the fine detail is not clear \citep{berger_etal1999,depontieu_etal2003}. There is bright loop
emission to the south that obscures part of the moss. The
XRT image shows that hot loops surround this area with core emission around the sunspots and
covering the moss. The detailed relationship between warm and hot emission in this region was
investigated in Paper 2.

The results of the detailed coalignment between {\it TRACE} and the SOT/SP map are shown in Figure \ref{fig3}. 
An overlay of 
the 171\,\AA\, image on the SP data is shown in the top panel. 
We identified the major
moss emission visually with a contour level $\sim$25\% of the brightest loop emission
in the 171\,\AA\, image. This is shown by the solid blue line.
Within this region, there 
is clearly bright loop emission. To exclude these structures, we 
drew another contour at the 45\% level (dashed blue line). The pixel coordinates of all points
inside the moss contour but outside the dashed region were recorded for all the images.
From here on when we refer to the moss region, this is the area we are referring to.

The magnetic field line inclination ($\theta$) and azimuth angle ($\phi$) in the region are
shown in the middle and lower panels of Figure \ref{fig3}
with the same contours overlaid. In spherical coordinates, the field line inclination is along the line-of-sight so that an angle of
0$^o$ is directed towards the observer, and an angle of 180$^o$ is directed away from the observer. This
will be relative to a radial field-line provided that the line-of-sight is normal to the surface, for example,
when an active region is at disk center. For the data shown, the active region was at $\sim$ 30$^o$ West of
disk center. 

The inclination angle in the figure is represented by the function 
\begin{equation}
\label{eq1}
f (\theta) = \left\{ \begin{array}{rl}
\theta &\mbox{ if $\theta < \pi/2$} \\ 
\pi-\theta &\mbox{ if $\theta \ge \pi/2$}
\end{array}\right.
\end{equation}
This image shows large values of the inclination as white, and smaller values as black.

It has been pointed out previously that the magnetic field line orientation 
can be almost vertical to the solar surface \citep{katsukawa&tsuneta_2005}.
The field rooted in the moss in this region appears to be only moderately inclined,
with strongly inclined field mostly around the edges of the moss or in the inter-moss lane.
It is also notable
that the inclination is coherent, i.e., there is no obvious mixing
of widely differing inclination angles. 

%%%%
The azimuth angle of the field is of course difficult to interpret because of the 180$^o$
ambiguity. For these data, the angle is measured from 0 to 180$^o$ from the Solar West 
position (RHS), but the value of the angle could be 180$^o$ in the opposite direction.
Therefore, following \citet{kubo_etal2007}, we show a $\cos 2\phi$ representation in
Figure \ref{fig3}.
This representation shows magnetic field oriented East-West as white, and 
field oriented North-South as black.

The strong moss emission is characterized by field oriented East-West in both polarities. The orientation
again seems coherent, with the field mostly changing direction around
the edges of the moss. The edges then, would be the places where most of 
the shear in the
magnetic field is located. The strongest shearing is close to the inter-moss region where
the field is predominantly oriented North-South. 

%%%%
\section{Modeling}
\label{model}
It has previously been noted that the large scale pattern of TRACE 171\,\AA\, brightness in the
moss evolves slowly, with dynamics of motion and variability mostly on small-scales \citep{berger_etal1999,tberger_etal1999}.
\citet{brooks_etal2008} noted that although there was fine scale variability in the magnetic
flux below the moss in the region they studied, the general pattern also evolved slowly. In this paper, we
assess the stability and 
evolution of the magnetic flux and vector field in AR 10960, but we also put our statements
on a quantitative basis by comparing the observed time-scales to 
typical loop cooling time-scales
computed from hydrodynamic simulations. Loop lengths vary throughout an active region, and 
since the radiative and conductive cooling times for a loop 
are dependent on the loop length, we first determined the distribution of 
lengths for loops relevant to
the moss region using potential field extrapolations. The real magnetic field in the solar atmosphere
is not likely to be current free (see e.g. \citealt{derosa_etal2009}), however, the results from \S \ref{characterB}
show that the magnetic field is sheared mostly in the inter-moss region and that the field in the moss
itself is unsheared with a small inclination angle. Since the moss is also unipolar, the field should
escape almost vertically to the solar surface since it has nothing to connect to locally even if the field
were non-potential. The potential field approximation, therefore, may be less problematic in the moss itself. In any
case, the extrapolation is
only used here to provide a realistic distribution of loop lengths.

\begin{figure}[ht]
\centering
\includegraphics[width=0.98\linewidth,viewport= 0 120 360 360]{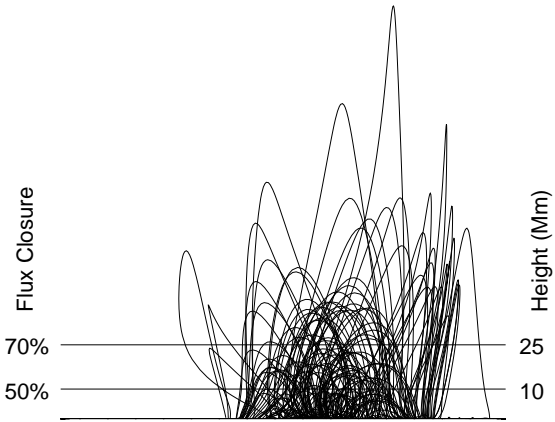}
\caption{Rotated view of the potential field extrapolation of 
Figure \ref{fig4} showing that, in contrast to the quiet Sun, most of the moss field connects to the corona. 
\label{fig5}}
\end{figure}
A magnetogram for extrapolation was prepared by weighting the SP magnetic field strength by the magnitude of the
Stokes V signal. The weighting determines the polarity of the flux. An area of 208$'' \times$160$''$
around the core of the active region was then extracted. Field lines were computed for every
pixel in the magnetogram with a field strength between 5 and 5000G. Field lines rooted outside the
moss region or that left the computational domain were then discarded. 
This resulted in a final
dataset of 23917 field lines. A subset of 218 
are shown in Figure \ref{fig4} overlaid on the magnetogram.

%%NEW%%
An interesting result from this simulation is that the connectivity of the magnetic field in the
moss is somewhat different than that of the quiet Sun. \citet{close_etal2003} showed that approximately 50\%
of the magnetic flux closes within 2.5Mm in the quiet Sun, with only 5--10\% extending to heights 
greater than 25Mm. As mentioned, the magnetic field in the moss has nothing to connect to locally in the
unipolar regions, so it can extend higher. Figure \ref{fig5} shows a side-view representation of our simulation
with the height extension and flux closure percentages indicated. 
Considering only the field lines rooted in $>$ 20G field
(approximately comparable to the simulations of \citet{close_etal2003}), we find that about half 
extend beyond 10Mm with a significant fraction (30\%) extending to 25Mm and above.
Many of the shortest field lines in our simulations are around the edges of the moss or 
crossing the inter-moss region. In these regions the potential field extrapolation is less likely
to accurately represent the real sheared vector field, so the fraction of moss field lines that 
extend to significant
heights is probably higher.
This suggests that the moss is different than the quiet Sun in that most of
the magnetized chromosphere is in fact connected to the corona.
As we have shown (\S \ref{characterB}), this is also consistent with SP observations that the 
departure from the radial field line direction 
is only moderate.

\begin{figure}[ht]
\centering
\includegraphics[width=0.98\linewidth]{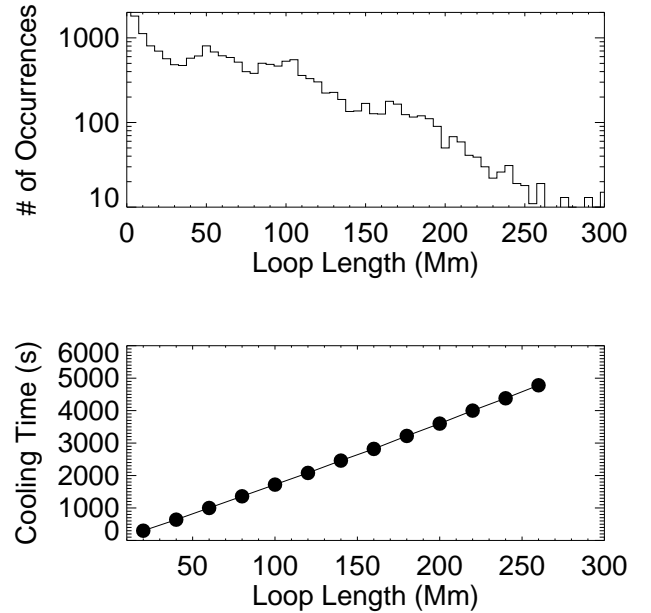}
\caption{Top Panel: Distribution of loop lengths from the potential field extrapolation for all closed field lines in the moss regions.
Bottom Panel: Loop cooling time versus length calculated by SOLFTM for the distribution of lengths in the top panel.
Definitions are in the text. 
\label{fig6}}
\end{figure}

The distribution of loop lengths for the full 23917 moss extrapolated field lines is shown in Figure \ref{fig6}
(upper panel). The distribution shows significant numbers of short ($<$ 30Mm) field lines and populations
in the range 30--80Mm and 80--120Mm. The maximum length is 337Mm.
To compute cooling times for the moss loops we prepared
a grid of lengths spanning a wide range up to 260Mm for input 
into the hydrodynamic code. For this simulation, we used the NRL Solar Flux Tube Model (SOLFTM) described
in detail by \citet{mariska_1987} and \citet{mariska_etal1989}. 
Each loop on the grid was allowed to cool from a starting equilibrium temperature of $\sim$ 4.8MK.
The apex temperature at each computational time step is computed by averaging over the loop top.
The loop is considered to have ``cooled'' when it reaches 1MK. This is close to the 
formation temperatures of \ion{Fe}{9} and \ion{Fe}{10}, % \citep{brooks_2010}, 
the spectral lines of which contribute significantly
to the {\it TRACE} 171\,\AA\, pass-band. The result of the simulation is shown in the lower panel
of Figure \ref{fig6}. The cooling times range from 300s for the shortest loops in the simulation, to 4800s
for the longest loops. A linear fit to the results gives a relationship
\begin{equation}
t_c = -127.7+18.7 L
\end{equation}
where $L$ is the loop length, and $t_c$ is the cooling time from 4.8 to 1MK.

\section{Variability and Evolution of Magnetic Field and Chromospheric Emission}
\label{variability}

\subsection{Magnetic Flux}
\label{mf_var}
\begin{figure*}
\centering
\includegraphics[width=0.98\linewidth]{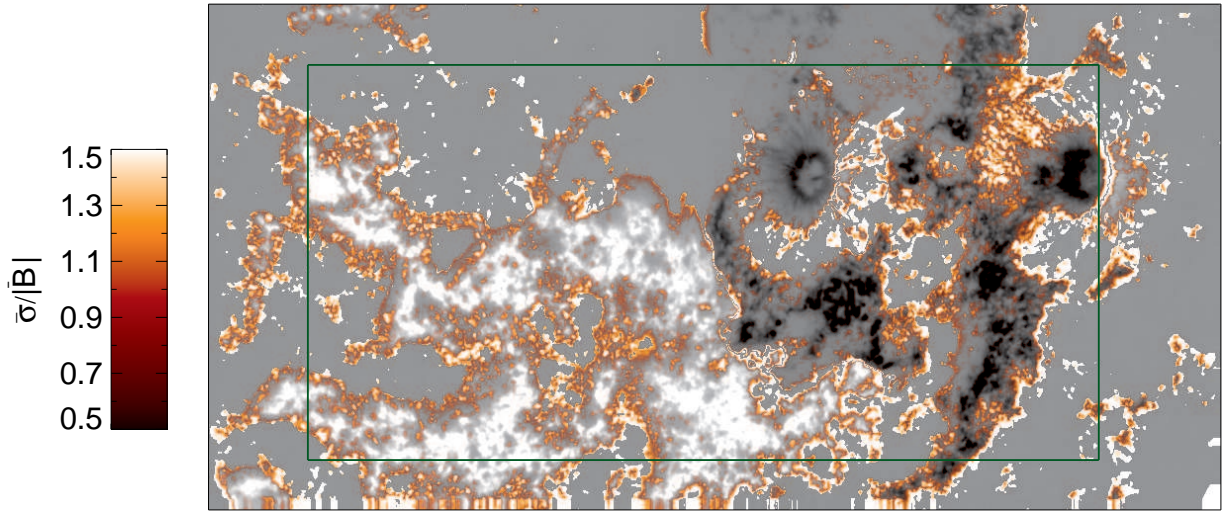}
\caption{Plot of $\bar{\sigma}/|\bar{B}|$ overlaid in red on $\bar{B}$ for each pixel in the complete time-series of SOT/FG data.
The solid green line shows the area used for the cross-correlation results of Figure \ref{fig8}. Only the points
where the magnetic flux is above the estimated statistical noise (16.6G) are shown.
A movie of the complete series of magnetograms is available in the electronic edition of the manuscript
as movie1.mpg.
\label{fig7}}
\end{figure*}

The time-series of FG magnetograms taken on June 09 was used
to assess the variability of the magnetic flux in AR 10960. 
For this global look, we co-registered the time-series using cross-correlation
on resampled (half dimension) data. We then 
calculated the average and standard deviation of the 
magnetic flux in each pixel over the whole time-series; $|\bar{B}|$ and $\bar{\sigma}$, respectively.
We then calculated the quantity $\bar{\sigma}/|\bar{B}|$, that gives a measure of the variability
in each pixel throughout the observations. This is over-plotted in red on $\bar{B}$ in Figure \ref{fig7}
so that the locations of high and low variability in the region can easily be distinguished.
Note that since $B$ is averaged over the whole time-series, $\bar{B}$ highlights the 
areas where the magnetic flux persists. Only the points were $\bar{B}$ is above the 
estimated statistical noise (see below) are plotted.

%%%%
As one would expect, the magnetic flux is persistent in and around the strongest field. It can
also be seen that the variability is highest away from these regions. In particular, the variability
is low in the moss and high around the edges. It is also high around the neutral line that passes through the inter-moss lane. In 
previous studies it has been shown that most transient brightening activity in the EUV occurs around the 
neutral line of an active region, and that the magnetic flux pattern is persistent around the bases
of high temperature loops \citep{brooks_etal2008}. This plot is consistent with that picture. 

The movie (movie1.mpg)
associated with Figure \ref{fig7} shows the spatial distribution of the magnetic flux evolving slowly.
To quantitatively asses this evolution we computed the linear Pearson cross-correlation coefficients, $r$,
between magnetograms separated by varying time-intervals. The Pearson coefficient is calculated by
dividing the covariance of two images by the product of their standard deviations, and was computed
here for the boxed green
area indicated in Figure \ref{fig7} using the IDL routine CORRELATE. 
Since each magnetogram is separated by only 30s, $r$ 
will be very high if the time-interval between them is small. As the separation is 
increased, the correlation will begin to break down because of the evolution of the spatial distribution
of the flux. The data are interrupted by {\it Hinode} night time every orbit, so the cross-correlation is 
made only between magnetograms taken in the same orbit.

\begin{figure*}
\centering
\includegraphics[width=0.80\linewidth]{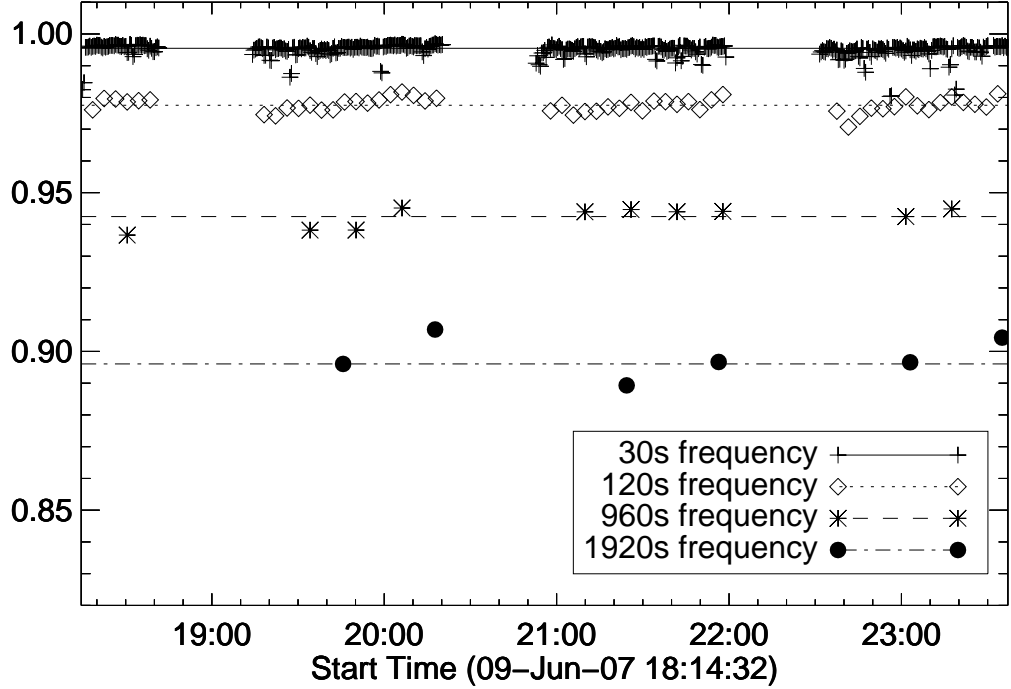}
\caption{Variation of the linear Pearson correlation coefficient between magnetograms separated by varying time-intervals as a function of time. The correlation is made between time-series with magnetograms taken at frequencies
of 30s, 120s, 960s, and 1920s. The last value is equivalent to the cooling time for a loop of length 110Mm.
\label{fig8}}
\end{figure*}

Figure \ref{fig8} shows the results. For magnetograms taken at 30s frequency $r$ is close to 1.00 for the
duration of the time-sequence, indicating a strong correlation between successive magnetograms as expected.
As the separation between magnetograms is increased to 120s $r$ falls, but is still maintained close to 0.98.
With a separation of 960s $r$ is close to 0.94, and with a separation of 1920s $r$ is maintained close to 0.90.
1920s is equivalent to the cooling time for a loop of length 110Mm in our hydrodynamic simulations. Since 85\% of the loops
in our field extrapolation
are shorter than this they cool from $\sim$ 5  to 1 MK on shorter time-scales.
This indicates that the spatial distribution
of the magnetic flux maintains a strong correlation for time-scales longer 
than the cooling times for
{\it at least} 85\% of the loops extrapolated from the moss regions. With the data at hand, it is not possible to assess the pattern
of magnetic flux over longer time-scales.

\begin{figure*}
\centering
\includegraphics[width=0.88\linewidth,viewport= 50 40 432 216]{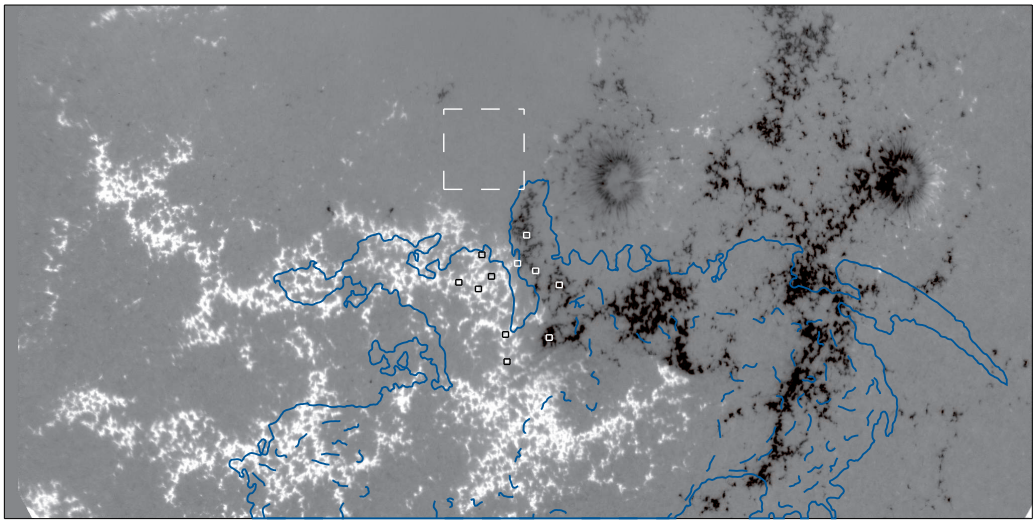}
\caption{FG \ion{Na}{1} 5896\,\AA\, magnetogram taken at 18:14:32UT on 2007, June 09. 
The large dashed box is the control region used for estimating the statistical noise
(see text).
Eleven small boxes are also overlaid and the variability of
the magnetic flux in these regions as a function of time is plotted in Figure \ref{fig10}.
\label{fig9}}
\end{figure*}

The variation of the magnetic flux in small areas as a function of time was then investigated. The analysis is
not sophisticated.
We used the full resolution data and extracted a 1000 pixel$^2$ area from each magnetogram. This time-series
was then co-registered via cross-correlation of successive magnetograms. Figure \ref{fig9}
shows one of the FG magnetograms in the time-series. Eleven small boxes are overplotted scattered throughout the positive
and negative polarity moss regions. One in the inter-moss region is also included.
The boxes are 2$'' \times$2$''$. This size was chosen because it is comparable to the spatial resolution
of EIS, which was measured to be close to 2$''$ in the laboratory pre-launch \citep{korendyke_etal2006}. 
It is about the same size as a low resolution MDI pixel. 

\begin{figure*}
\centering
\includegraphics[width=0.20\linewidth]{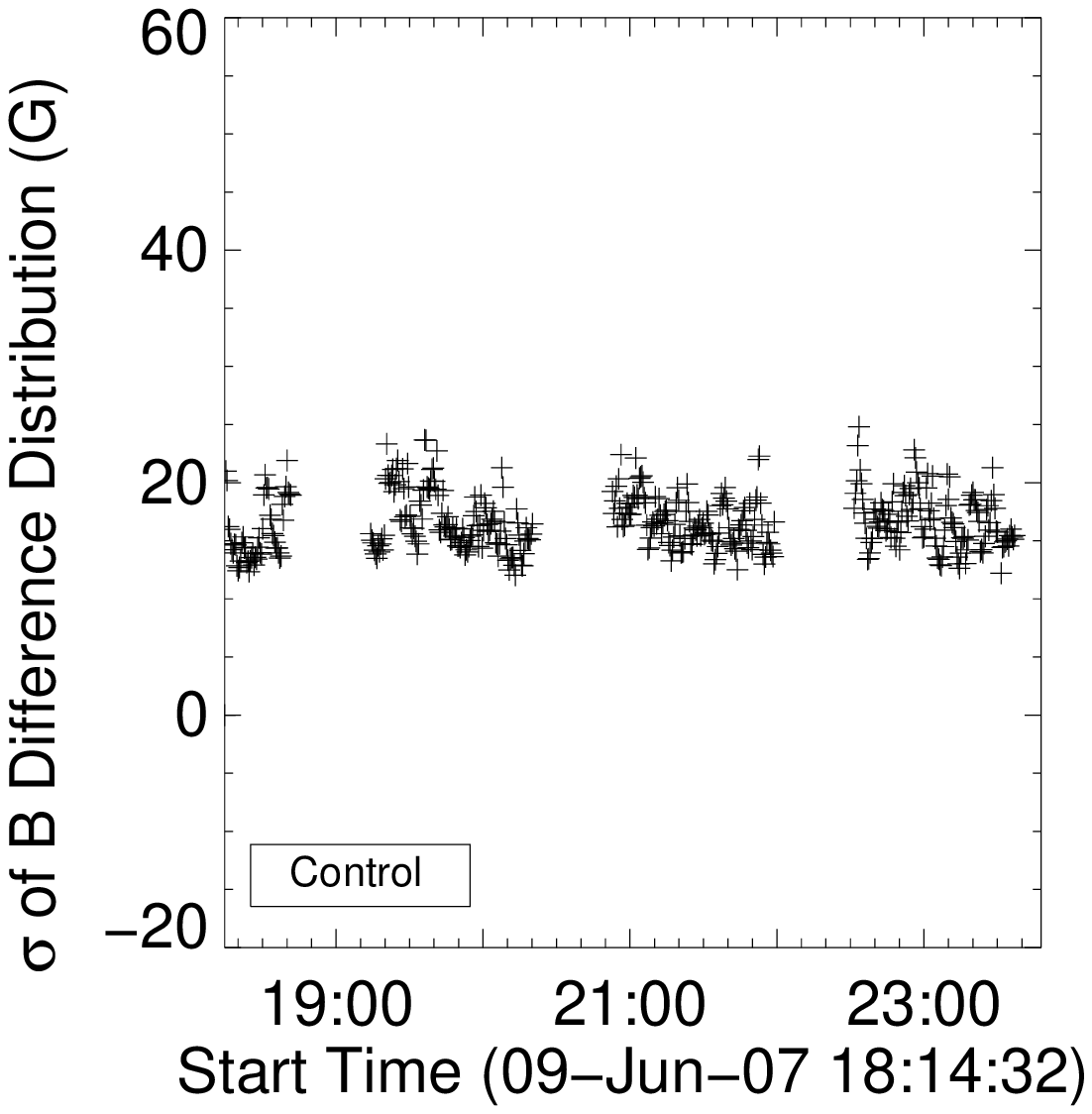}
\includegraphics[width=0.20\linewidth]{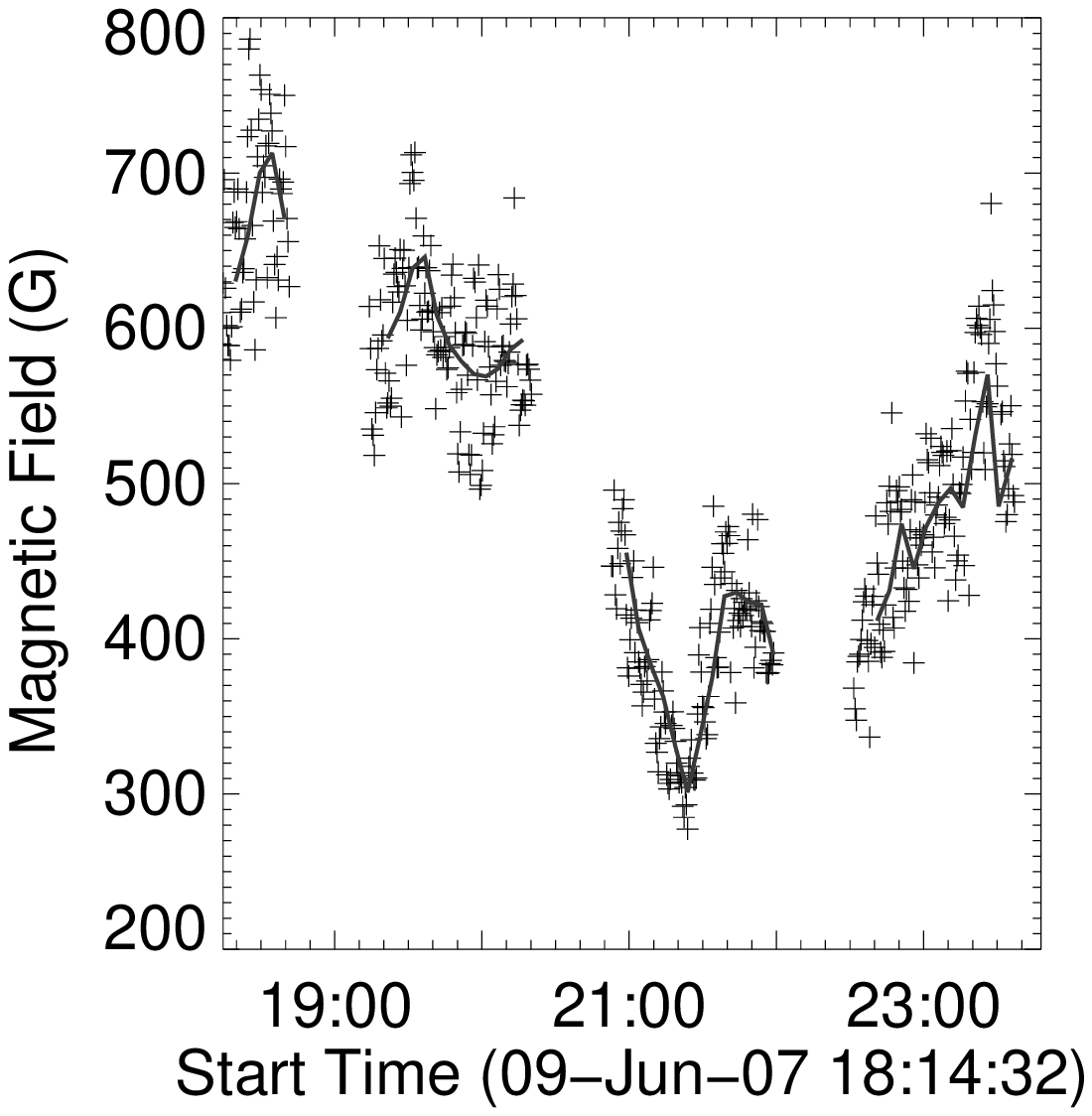}
\includegraphics[width=0.20\linewidth]{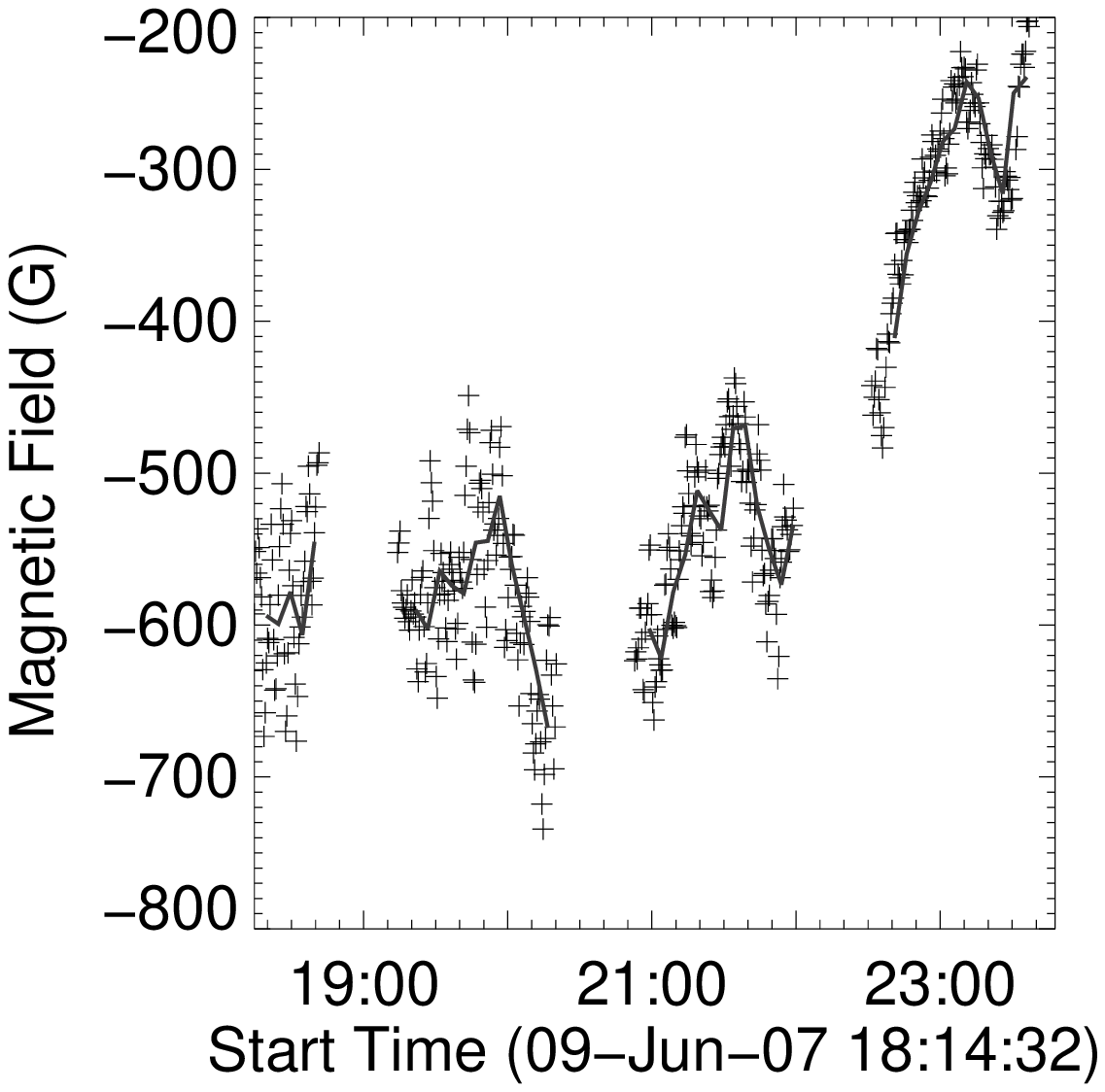}
\includegraphics[width=0.20\linewidth]{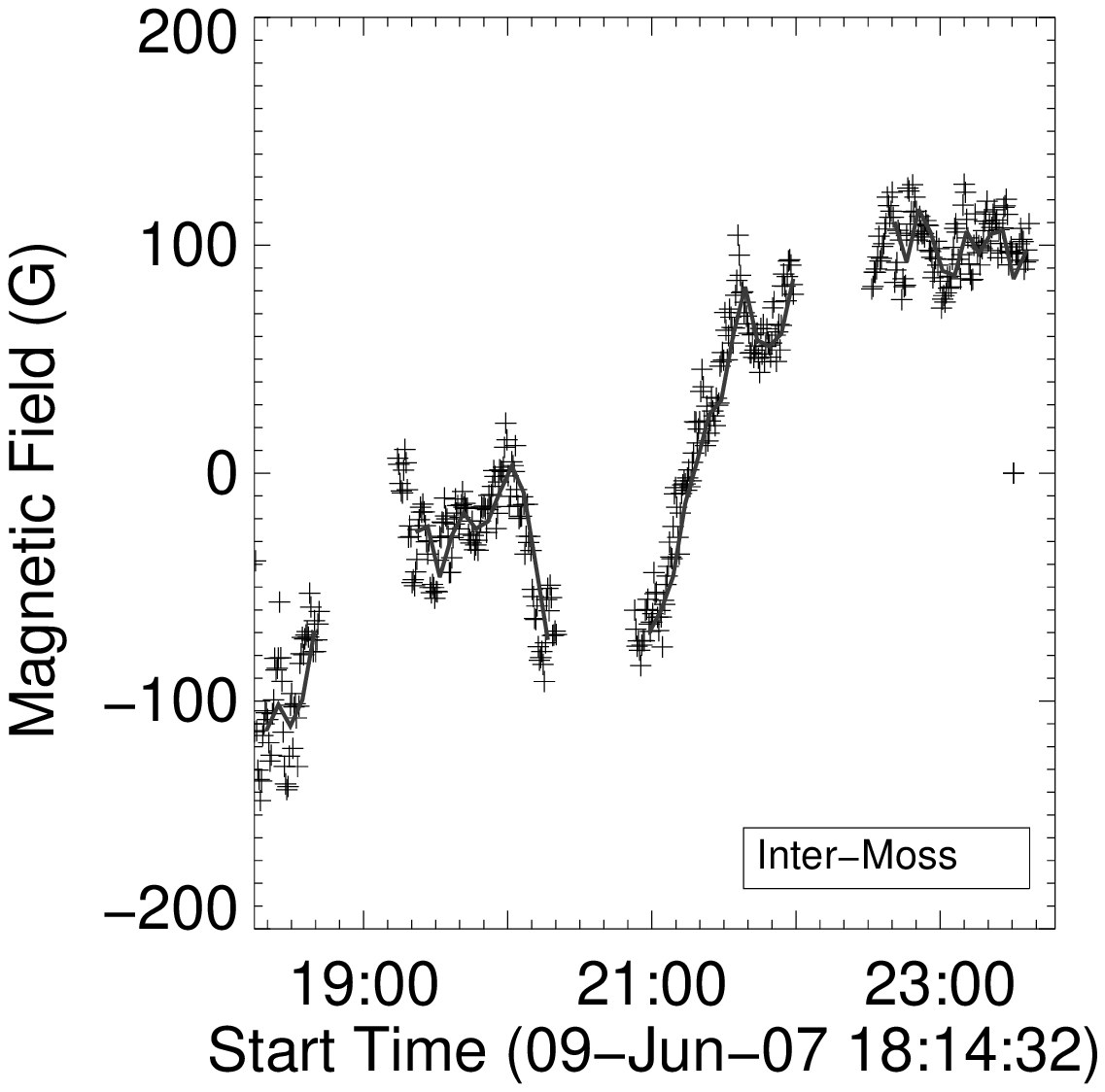}
\includegraphics[width=0.20\linewidth]{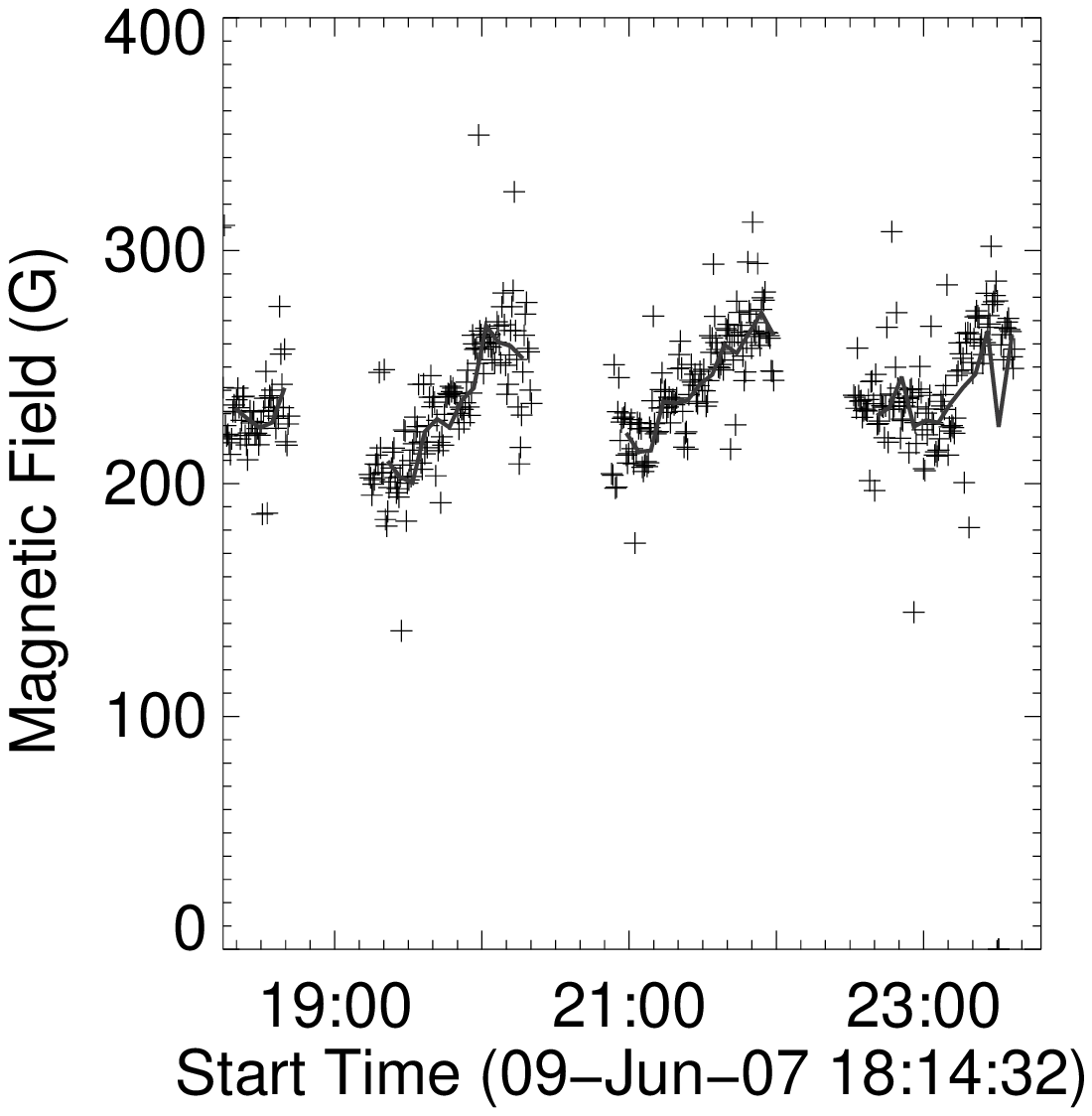}
\includegraphics[width=0.20\linewidth]{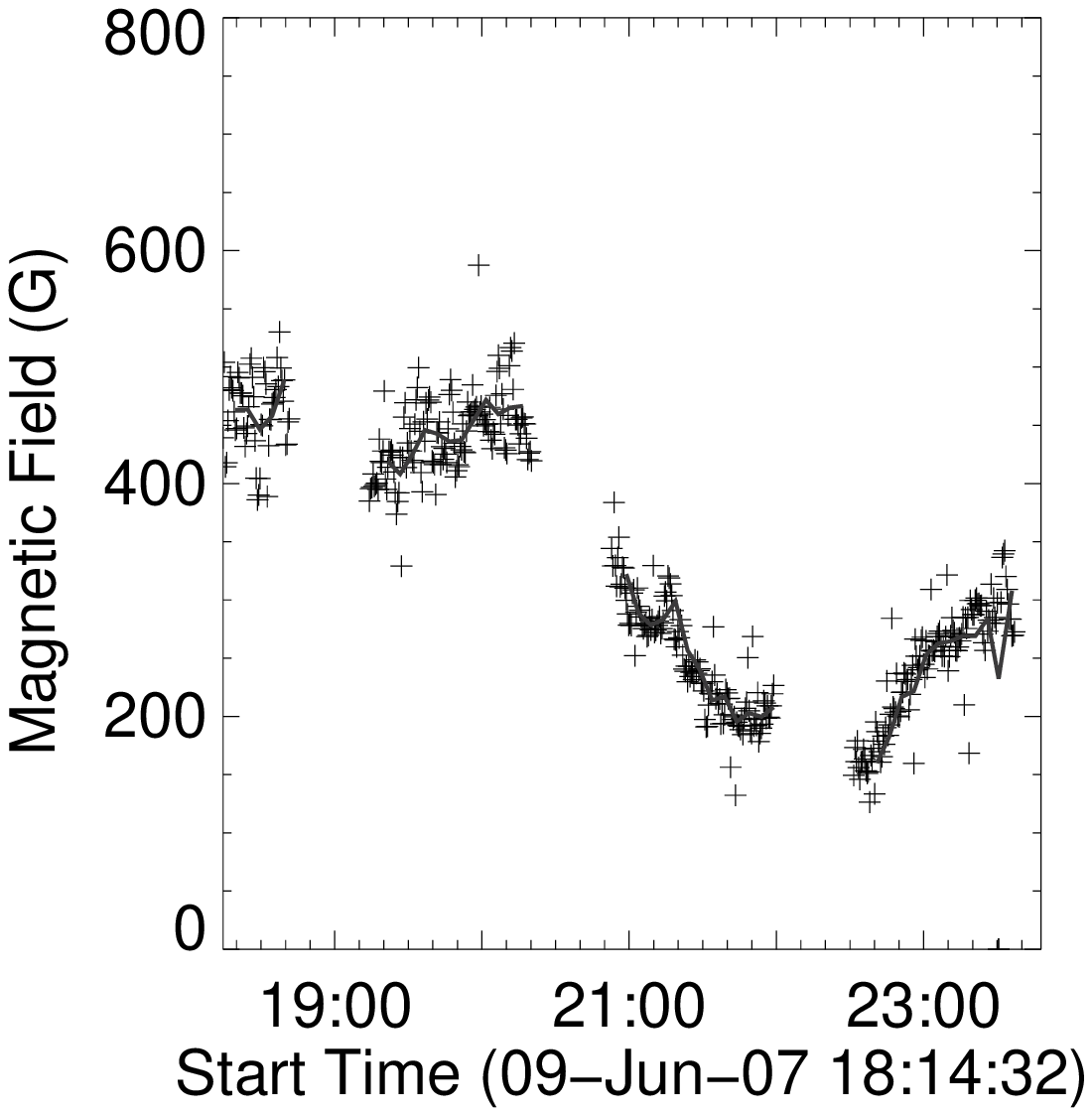}
\includegraphics[width=0.20\linewidth]{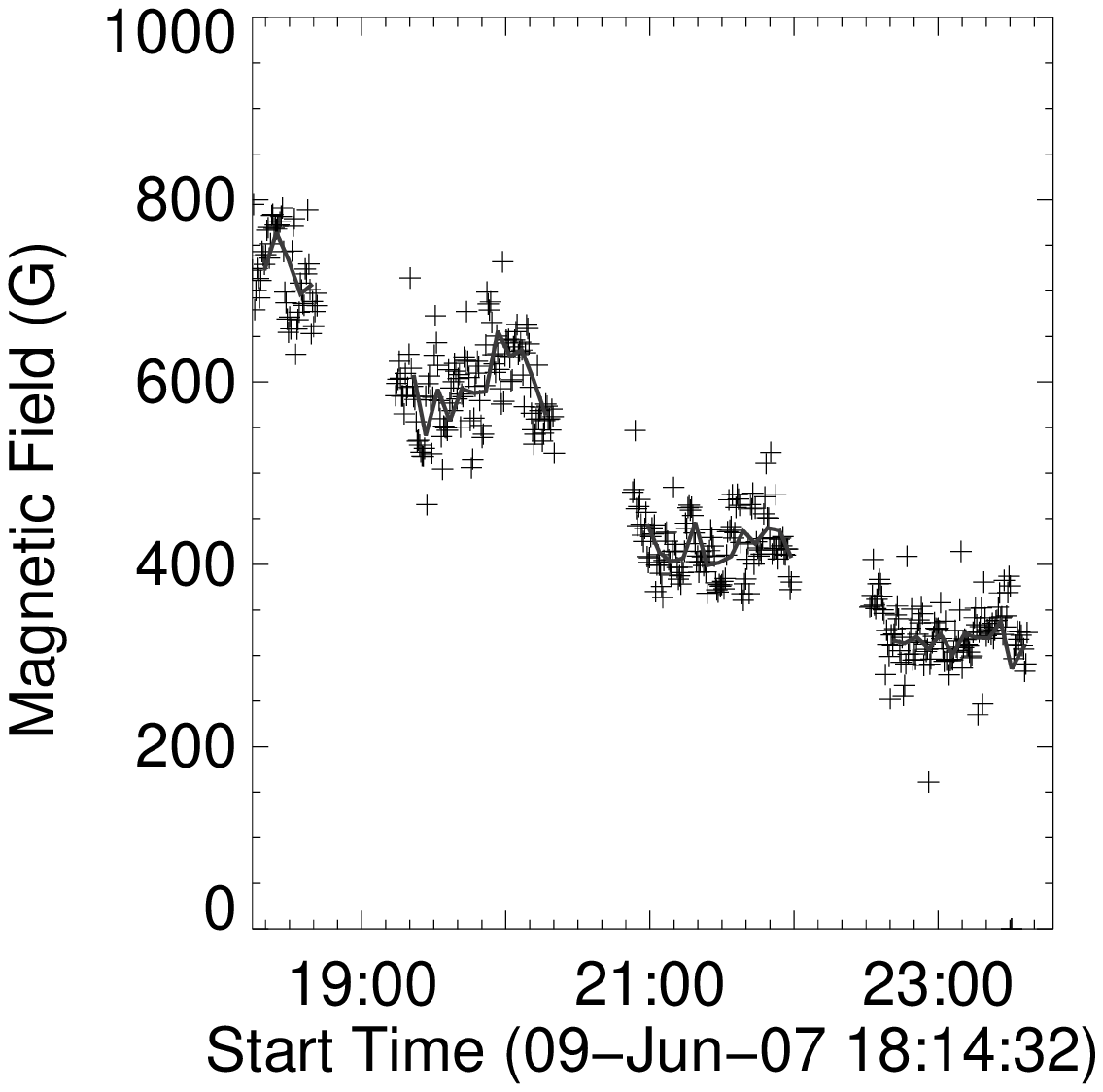}
\includegraphics[width=0.20\linewidth]{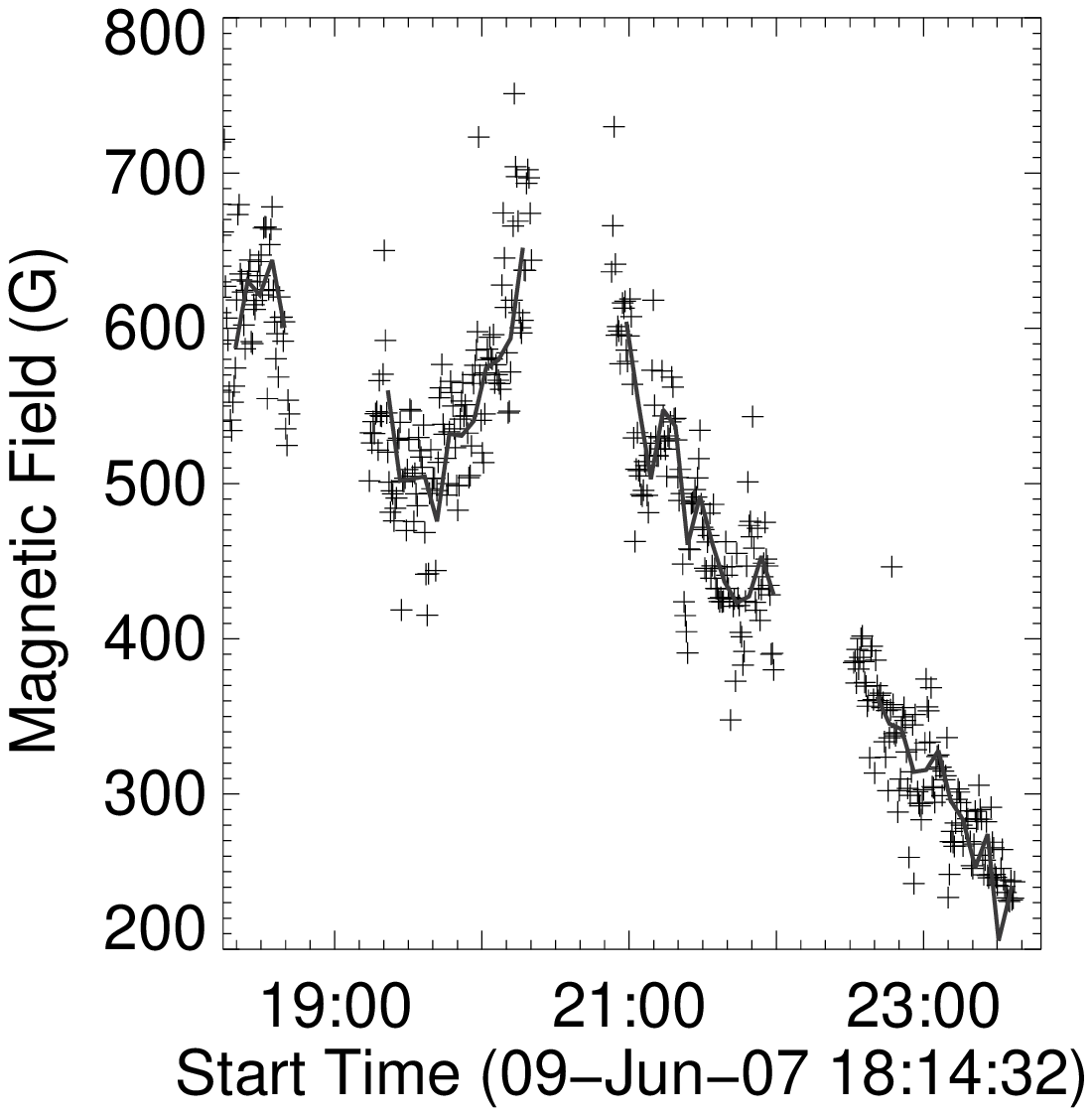}
\includegraphics[width=0.20\linewidth]{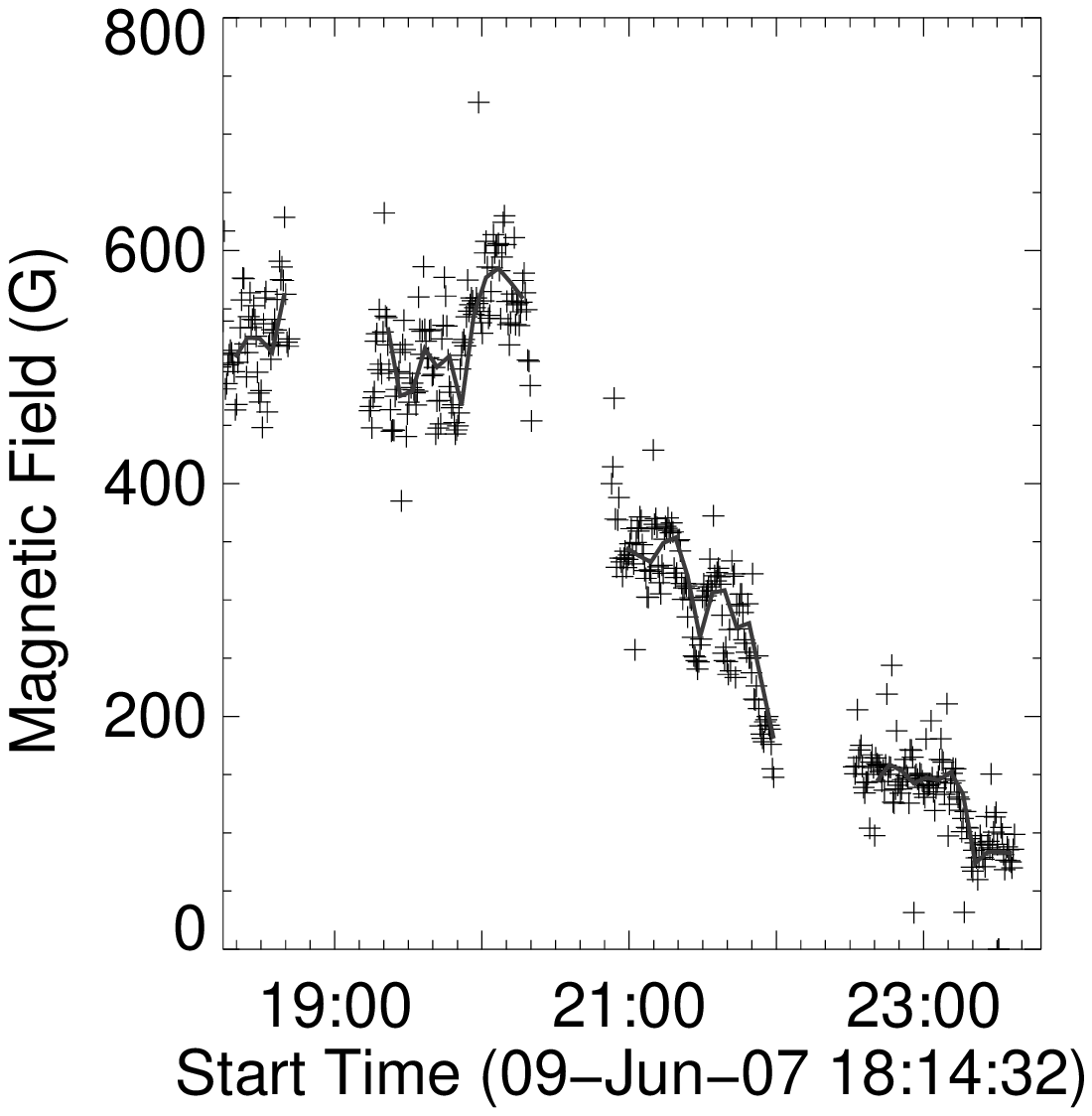}
\includegraphics[width=0.20\linewidth]{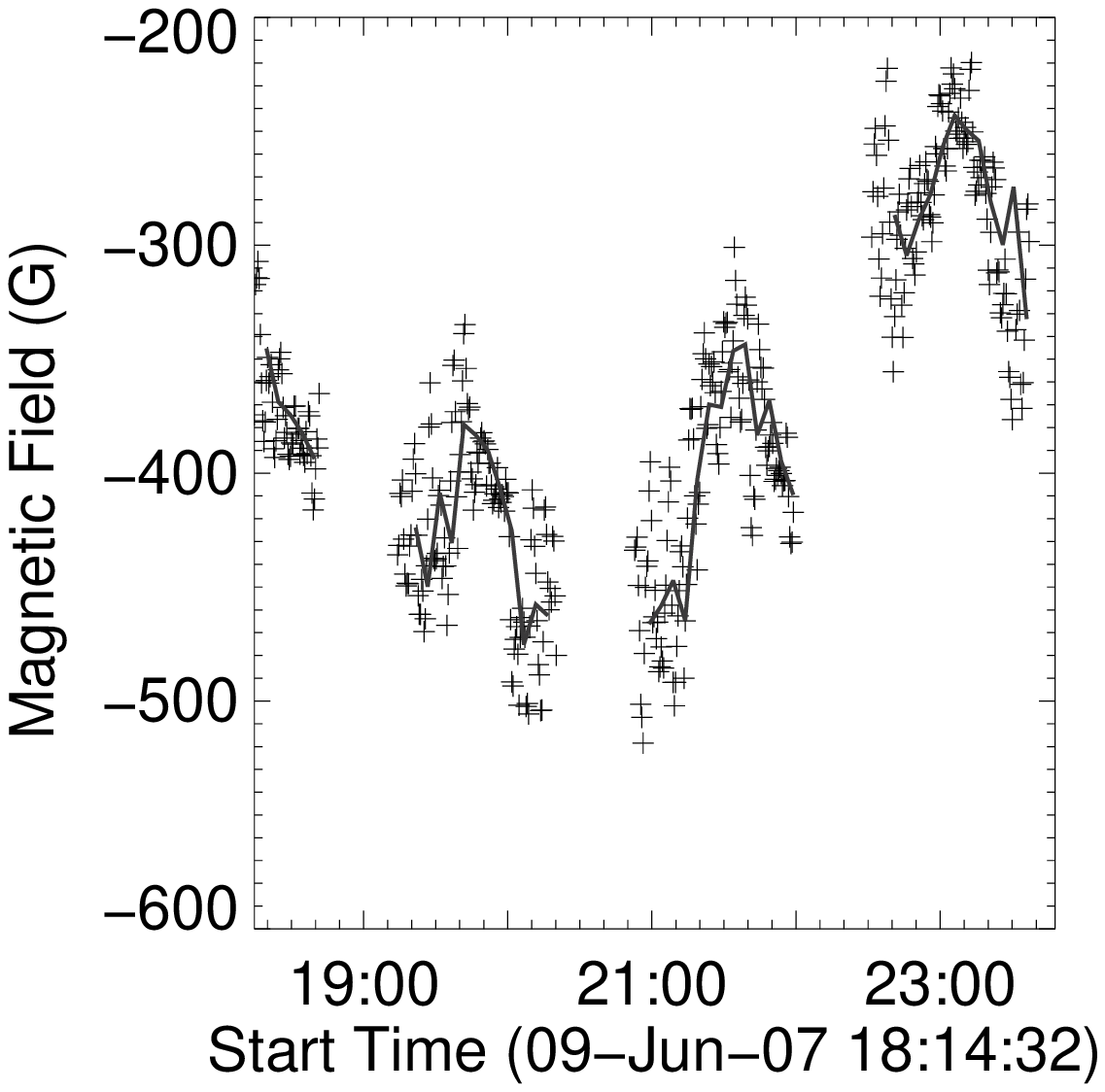}
\includegraphics[width=0.20\linewidth]{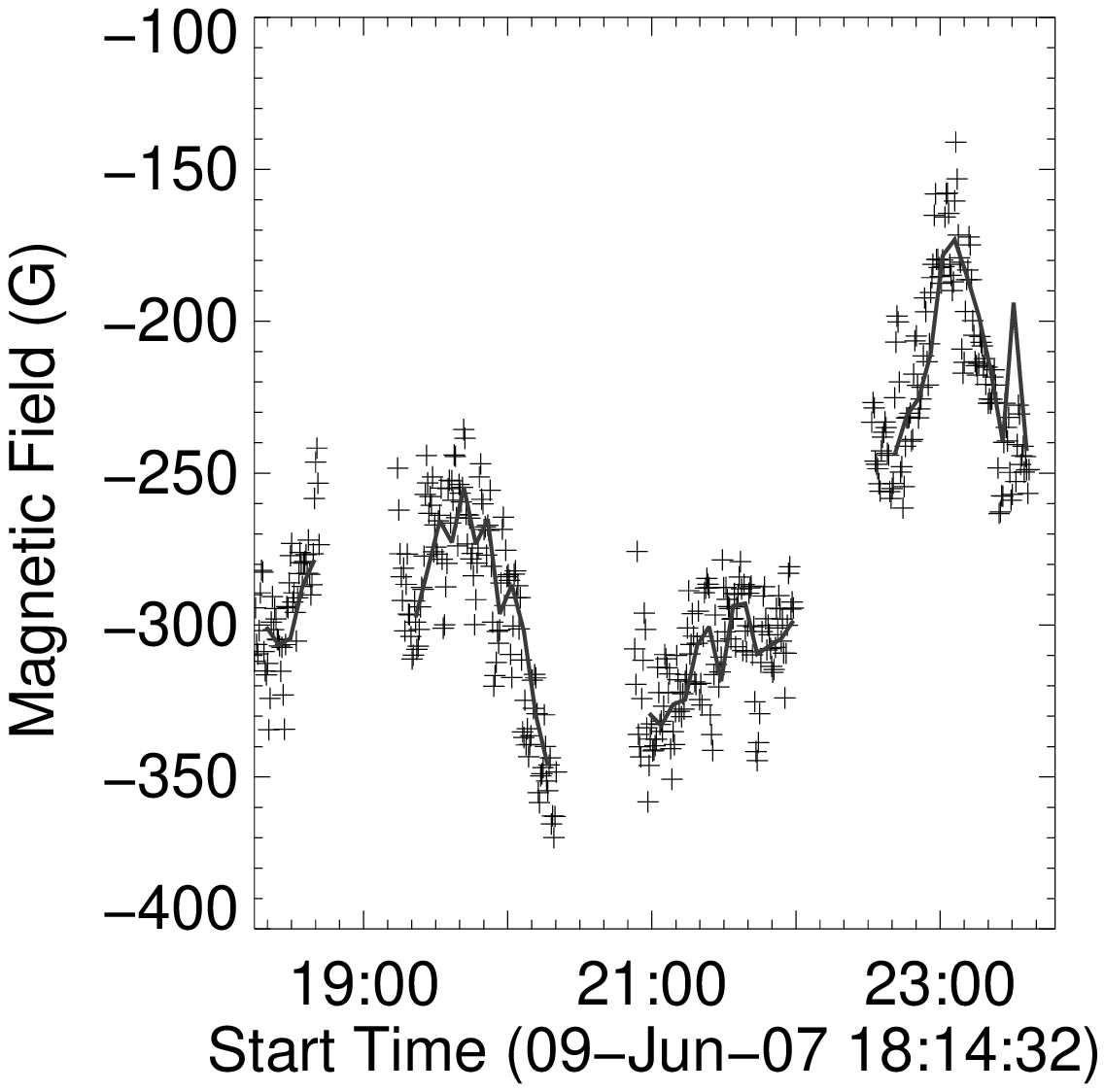}
\includegraphics[width=0.20\linewidth]{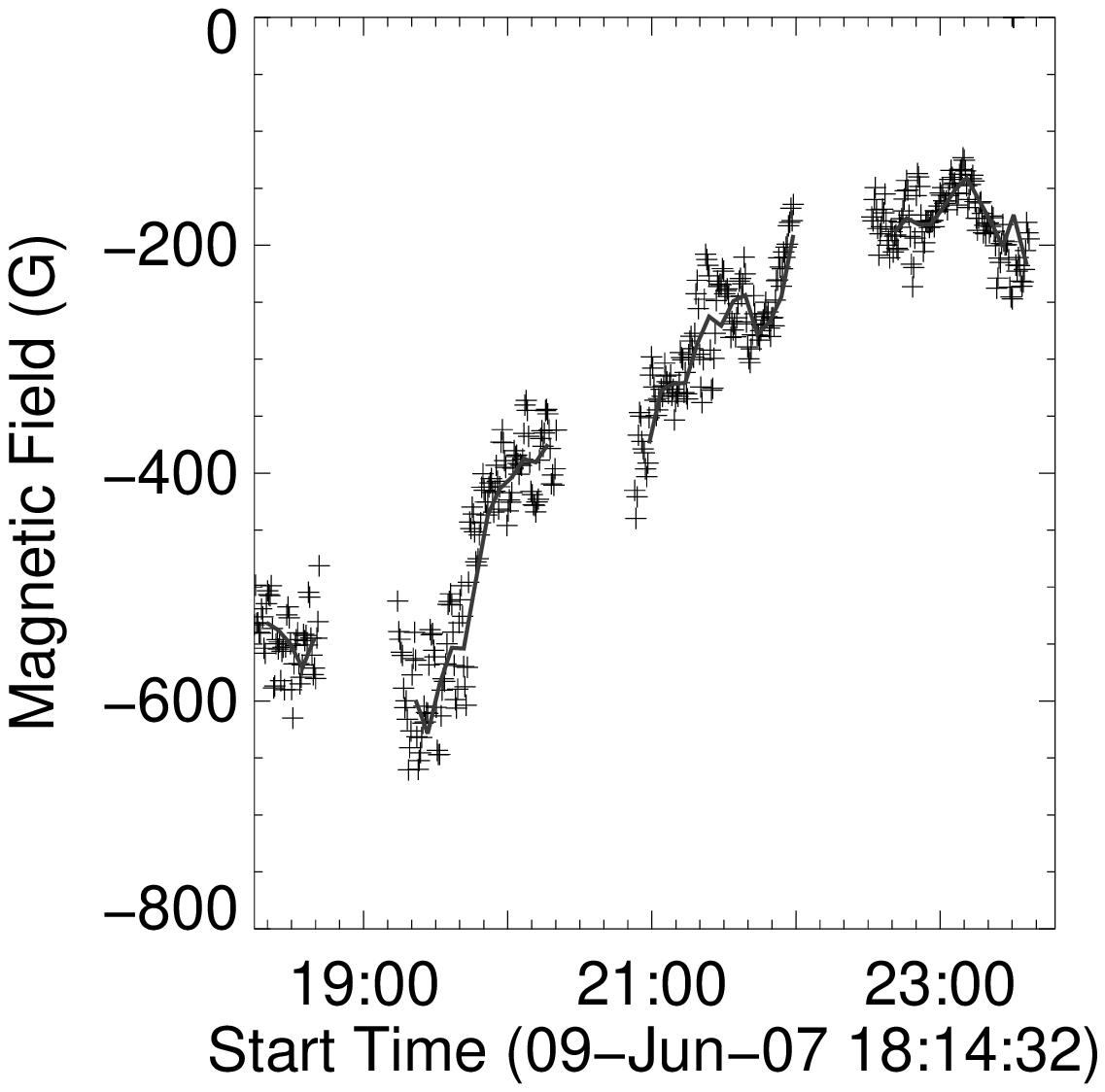}
\caption{Variation of moss magnetic field - flux per unit area - in Gauss (G) as a function of time. Top left: standard deviation
of histograms of the pixel-to-pixel difference between successive magnetograms in the control box shown in Figure \ref{fig9}.
Other panels: evolution of the magnetic field averaged over the small boxes in Figure \ref{fig9}. The crosses show the 
individual points, and the thick solid lines show detrended 5 mins averaged curves to de-emphasize the scatter.
\label{fig10}}
\end{figure*}

In order to understand whether the variations in small areas are significant or not, we need to estimate the
statistical uncertainty. Since the FG does not do a full magnetic vector inversion we cannot quantify the
various sources of noise. Therefore, we adopted the following approximate method.
We selected a relatively quiet region, shown by the large box in Fig \ref{fig9}, and
formed 
histograms of the difference in magnetic flux per unit area between successive magnetograms in the time-series. These histograms
were then fitted with Gaussian functions and the standard deviation ($\sigma$) measured. The standard deviations are
plotted as a function of time in the upper left panel of Fig \ref{fig10}. The average value is 16.6G with $\sim$
15\% variation in the standard deviations as a function of time. 
We adopt this average value as our estimate of the statistical noise.
There may be signals below this value, but they would be difficult to reliably distinguish from noise, with the
caveat that the quiet region selected is in an active region so is only `relatively' quiet. It is likely
that the true statistical noise level is lower. This estimate should be taken into account when considering
the results for the moss and inter-moss regions below.

Figure \ref{fig10} shows the results of this analysis. The interpretation is complicated because it is difficult to
distinguish changes in the magnetic flux due to features evolving, or moving in or out of the field of view. 
As noted
by \citet{brooks_etal2008}, the constancy of the spatial distribution of magnetic flux in an active region can
be preserved by magnetic features moving along similar paths, i.e., the field is dynamic, but the pattern is maintained.
We are interested, of course, in the evolution of features, but also want to eliminate cases where we end up
tracking something else.
We have therefore made an effort to select regions where obvious motions in or out of the box are reduced. The movie (movie2.mpg)
associated with Figure \ref{fig9} allows
the reader to independently judge this analysis.

The positive polarity moss box in the top row of Figure \ref{fig10} shows an average magnetic field of 
510G and around 22\%
variation over the duration of the observations ($>$ 5 hr). This is clearly much longer
than the theoretical cooling times. The negative polarity moss box in the top row
shows a comparable average magnetic field of 
-490G and $\sim$ 27\%
variation during the observations. The majority
of the boxes in Figure \ref{fig10} show fluctuations around this level of 
15--30\%.
These values for the moss 
are a little higher than the results found for coronal intensities and velocities
in papers 1 and 2. A few cases show larger variations at the 
40--50\% level, but it is clear that this also 
reflects the fact that the magnetic flux is evolving slowly.
Note, for example, the box in the lower right panel of Figure \ref{fig10}. This box shows a variation of
$\sim$45\% during the observing period. This is a result, however, of a slow evolution from around
-550G at 18:30UT to around -200G at 23UT
rather than a sudden change. In fact, even in this box, a less than 
30\%
variability is maintained if we consider just the first, or last, 3 hours of observations. If we consider
only the first hour of the observations 
the magnetic flux in all of the boxes varies by less than 10\%.
This is still considerably longer than most of the simulated loop
cooling times: only loops longer than $\sim$200Mm persist for an hour, and only about 2\% of the loops
in our simulation are this long.

%%%%
%%%%
The inter-moss region 
shows an average magnetic field of 15G and the highest variation of all the boxes ($\sim$ 60\%)
This is consistent with our expectations since the inter-moss region was already identified as being more
variable in Figure \ref{fig7}. As pointed out earlier, transient EUV brightenings have been found to be
preferentially located 
around the neutral lines in other active regions. The inter-moss region also seems to be where the majority
of the flares and transients occur in this region (see, e.g., the {\it TRACE} movies in papers 1 and 2).

Some areas of the moss do show evidence of dynamic changes in magnetic flux similar to that in the inter-moss
box. 
Also, it is unclear how significant 15--30\% variations in moss magnetic flux are over time-scales of
several hours. Variations 
comparable in magnitude do occur on shorter time-scales (tens of mins). These changes are
on the order of 100--200G and are statistically significant. 
Therefore, they could be taken as evidence for low-frequency impulsive heating. 
Note that changes on 20--30 mins time-scales correspond to the cooling times for loops in the 70--100Mm range.
We find, however, that 
only a small fraction ($\sim$10\%) of our simulated moss loops would be expected to cool on 
these time-scales. This suggests that such heating is not significant here, and may also
be consistent with the lack of ``warm" EUV loops cooling in the core of this region. 

The 100-200G magnetic flux changes correspond to $\sim$ 10$^{18}$Mx in the boxed regions, and would be equivalent to smaller than micro-flare
size energy releases in the corona. Further
work is needed to see if these could be related to, or signatures of, the flare-like intensity
variations intermittently seen in hot X-ray core loops \citep{warren_etal2007} or hot transient events \citep{shimizu_1995}.
Here we focus on determining whether such variations show evidence of propogating energy into the chromosphere. 

\subsection{\ion{Ca}{2} H Intensity}
\label{ca_var}
\begin{figure}[ht]
\centering
\includegraphics[width=0.70\linewidth,viewport= 40 40 360 360]{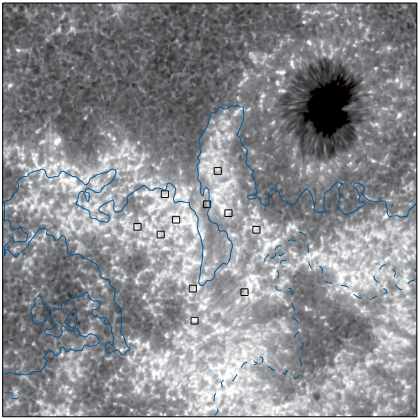}
\caption{FG \ion{Ca}{2} 3896\,\AA\, magnetogram taken at 18:14:54UT on 2007, June 09. 
Eleven small boxes are overlaid and the variability of
the \ion{Ca}{2} intensity in these regions as a function of time is plotted in Figure \ref{fig12}.
The moss contours from Figure \ref{fig3} are overplotted in blue.
\label{fig11}}
\end{figure}

The variation of chromospheric emission in the \ion{Ca}{2} 3896\,\AA\, line as a function of time was investigated using 
the time-series 
of images taken on June 09. The time-series was co-registered to the first image by cross-correlating
successive images. Figure \ref{fig11} shows an example image taken at 18:14:55UT with the same 
eleven small boxes and large control box as in Figure \ref{fig9} overlaid. The placement of these boxes was made by determining the inter-FG
offsets and scalings by co-registering this \ion{Ca}{2} image with a near-simultaneous \ion{Na}{1} D magnetogram. This 
was achieved by cross-correlating a common area from the two images and re-sampling the magnetogram to the higher
resolution of the \ion{Ca}{2} image.

\begin{figure*}
\centering
\includegraphics[width=0.20\linewidth]{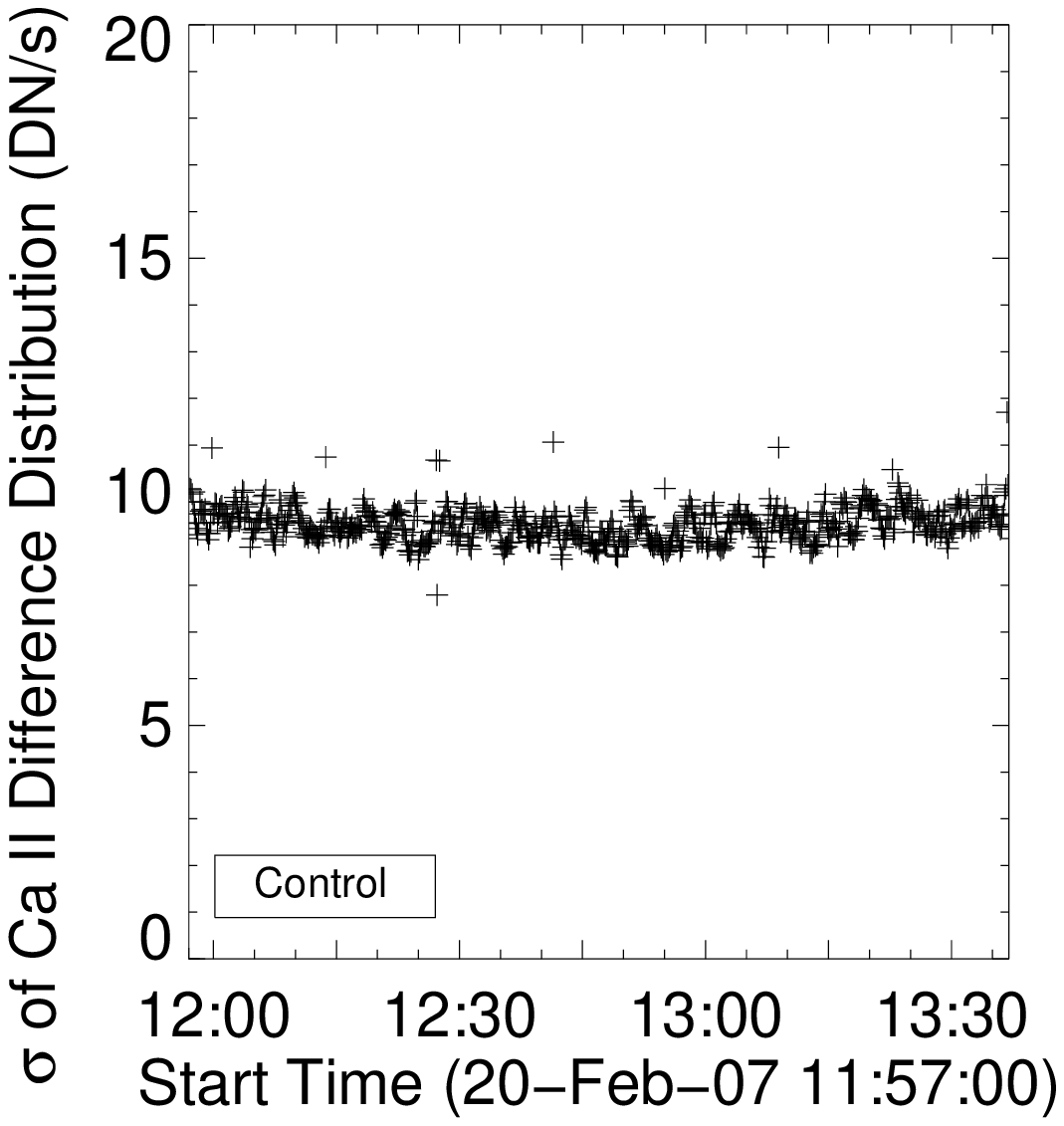}
\includegraphics[width=0.20\linewidth]{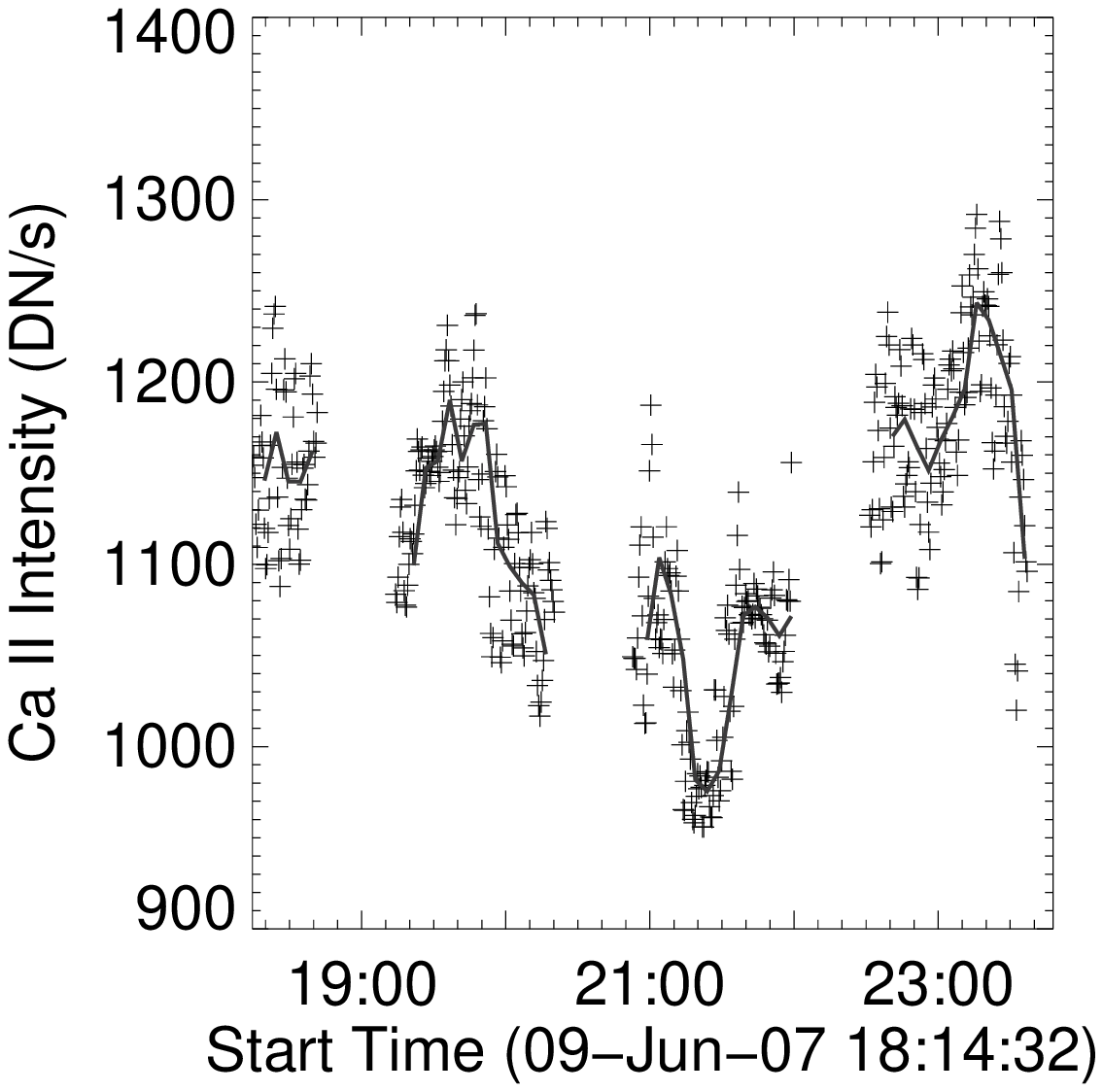}
\includegraphics[width=0.20\linewidth]{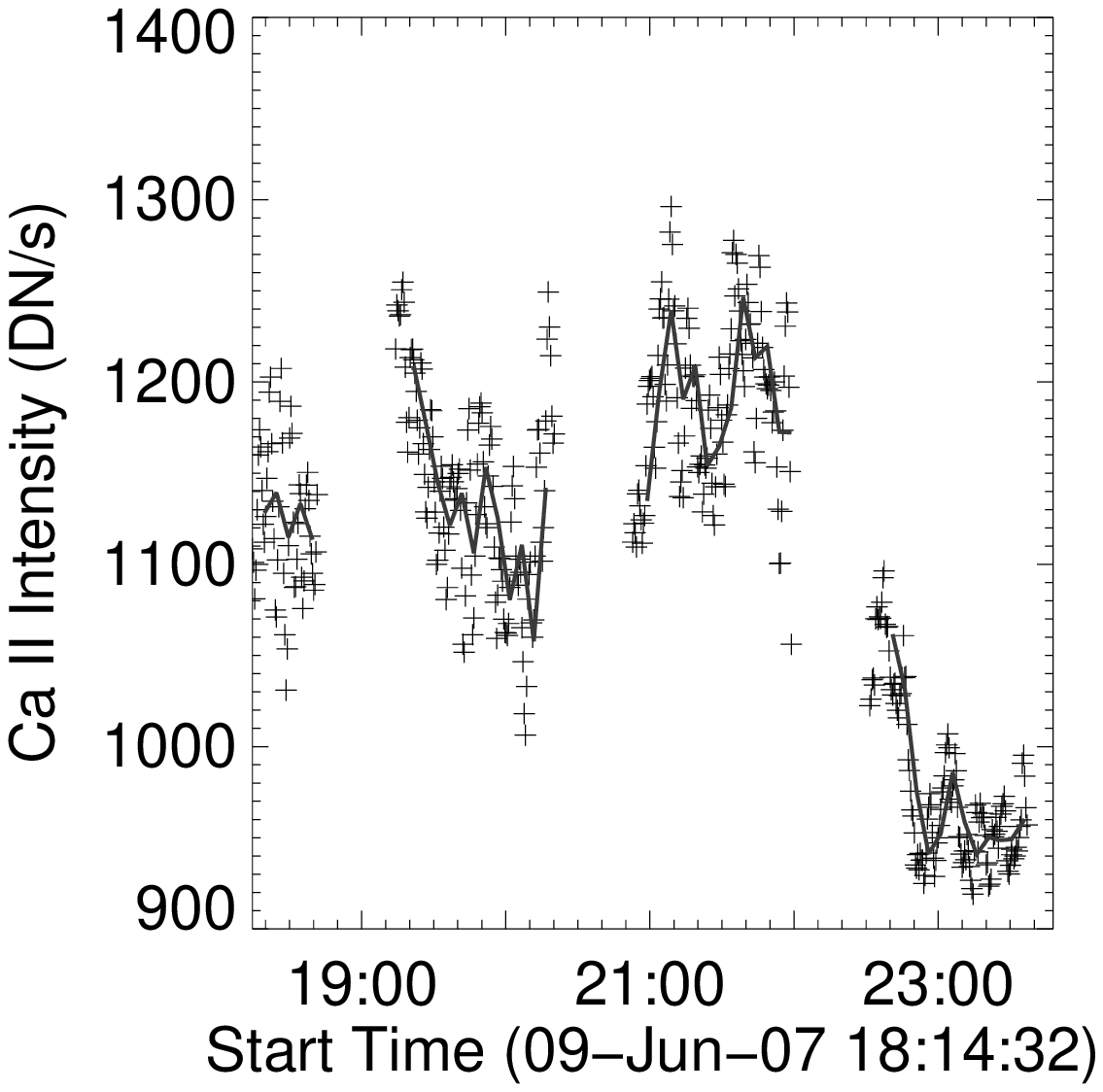}
\includegraphics[width=0.20\linewidth]{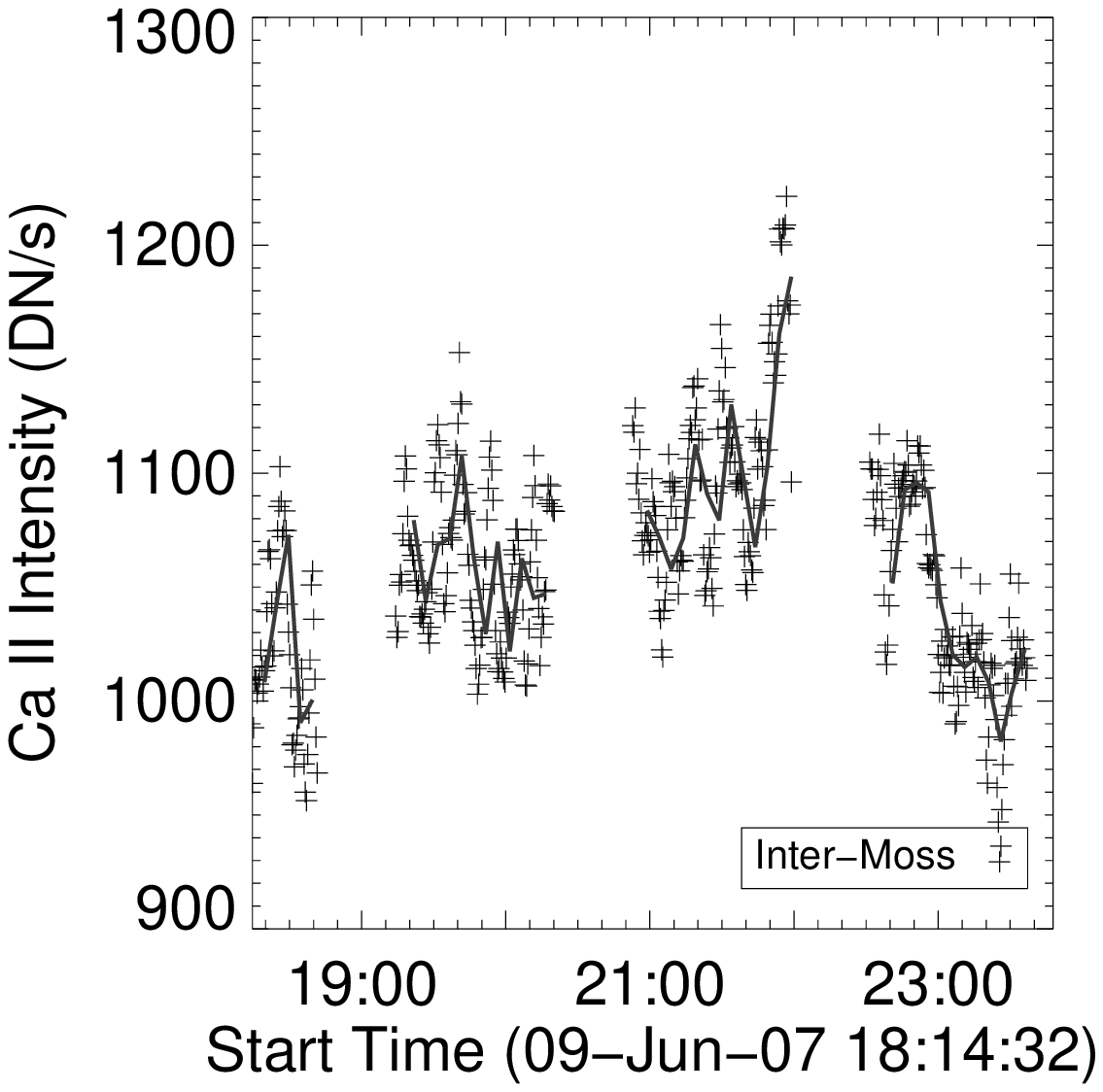}
\includegraphics[width=0.20\linewidth]{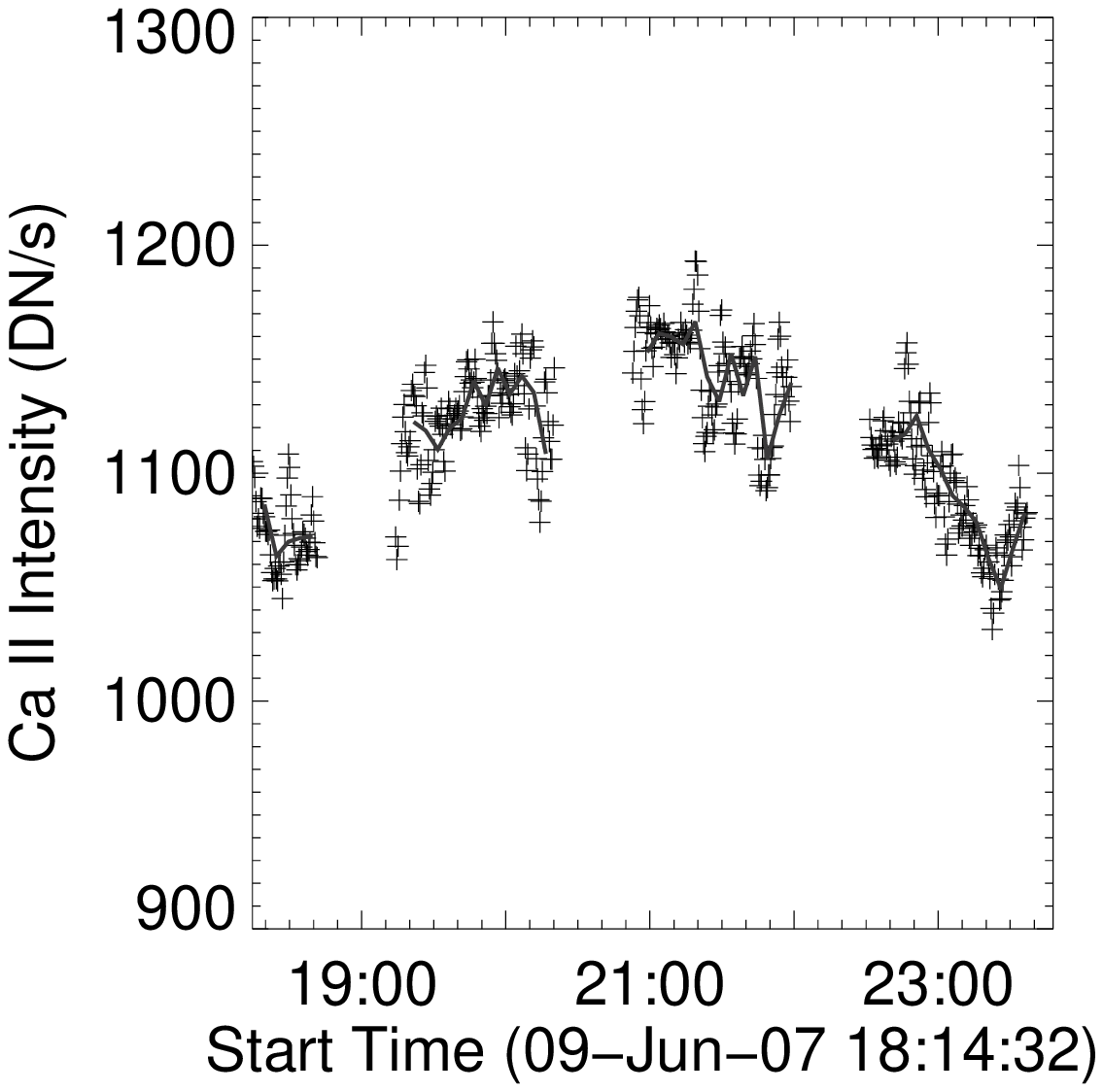}
\includegraphics[width=0.20\linewidth]{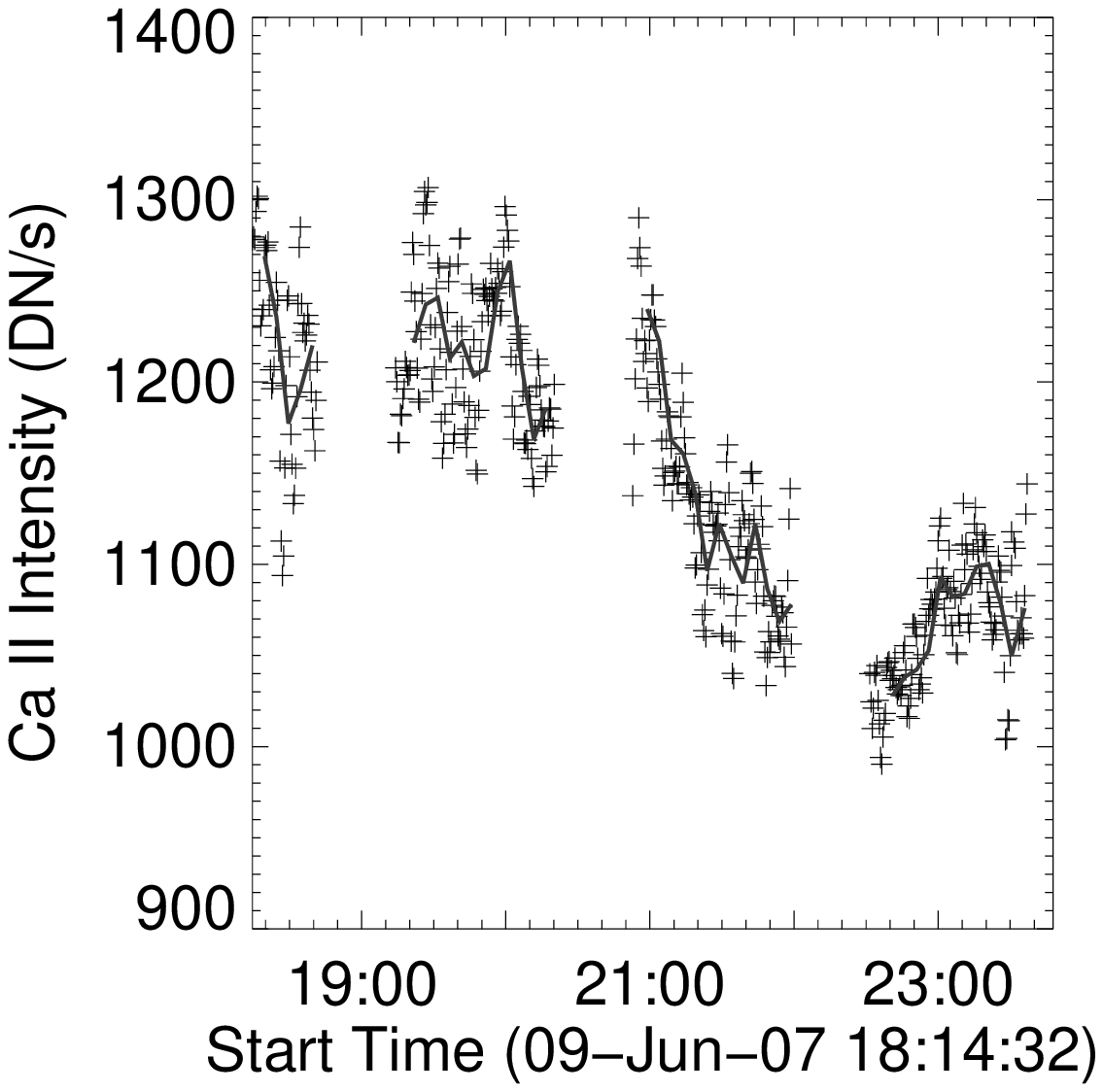}
\includegraphics[width=0.20\linewidth]{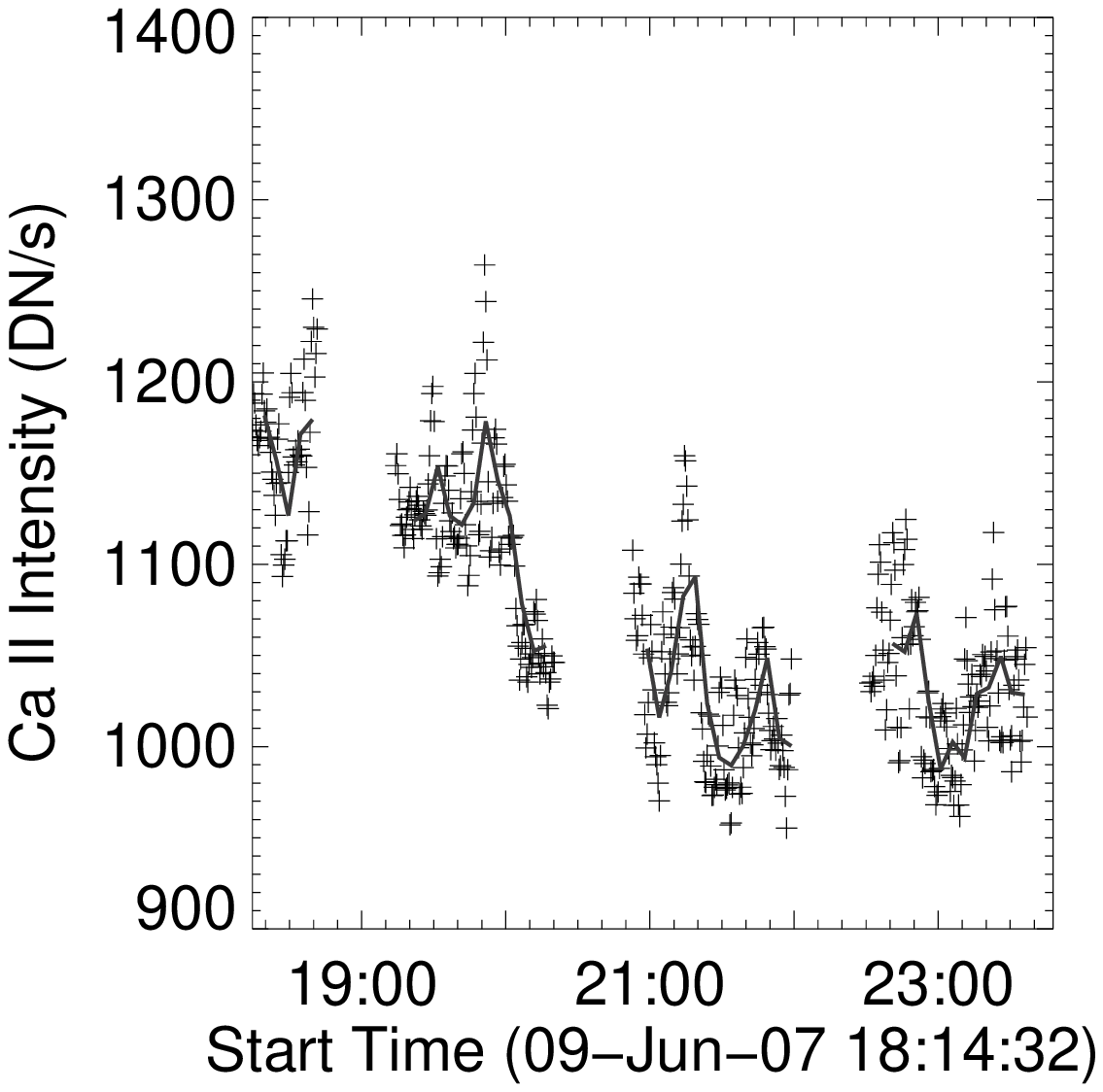}
\includegraphics[width=0.20\linewidth]{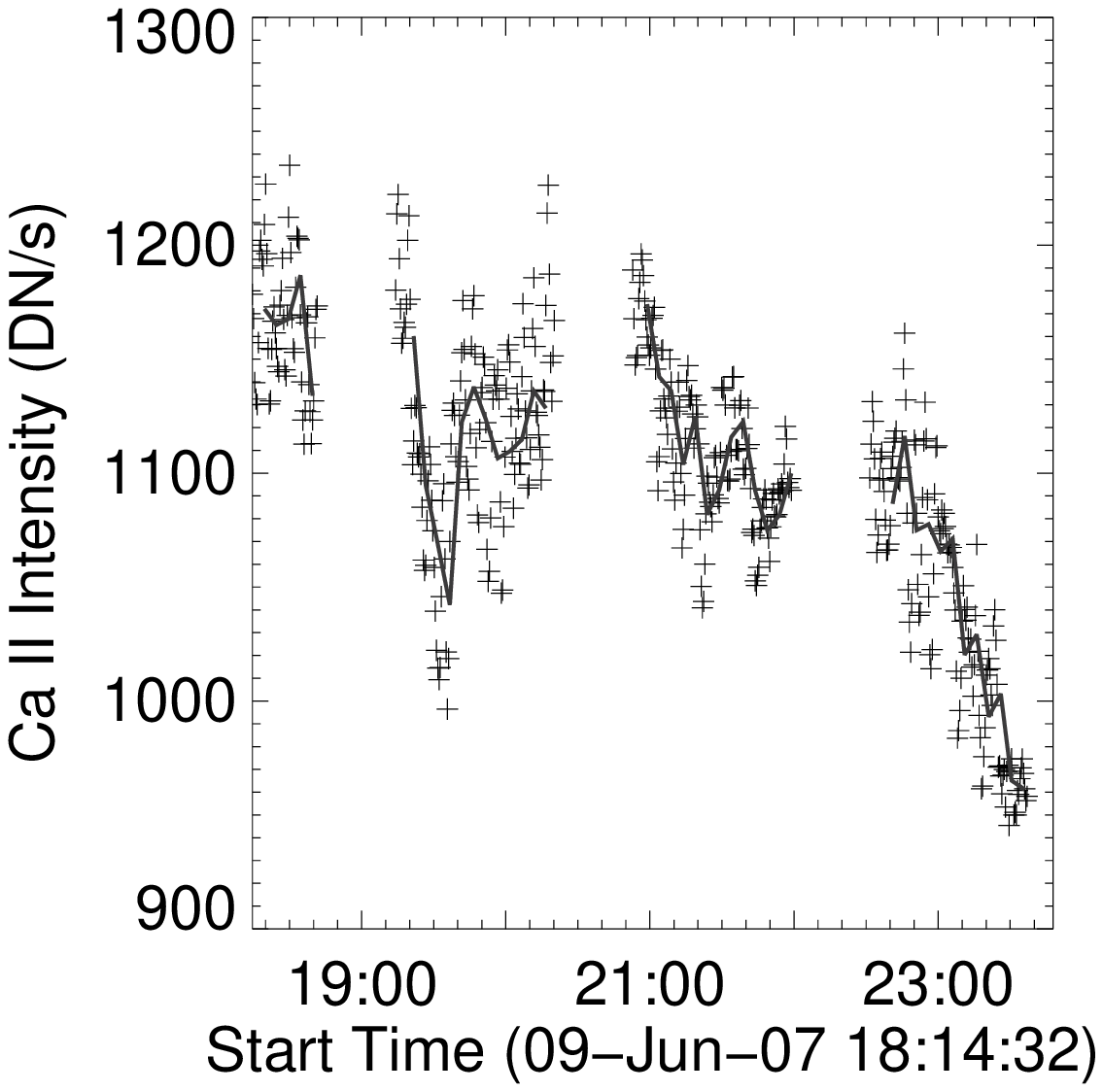}
\includegraphics[width=0.20\linewidth]{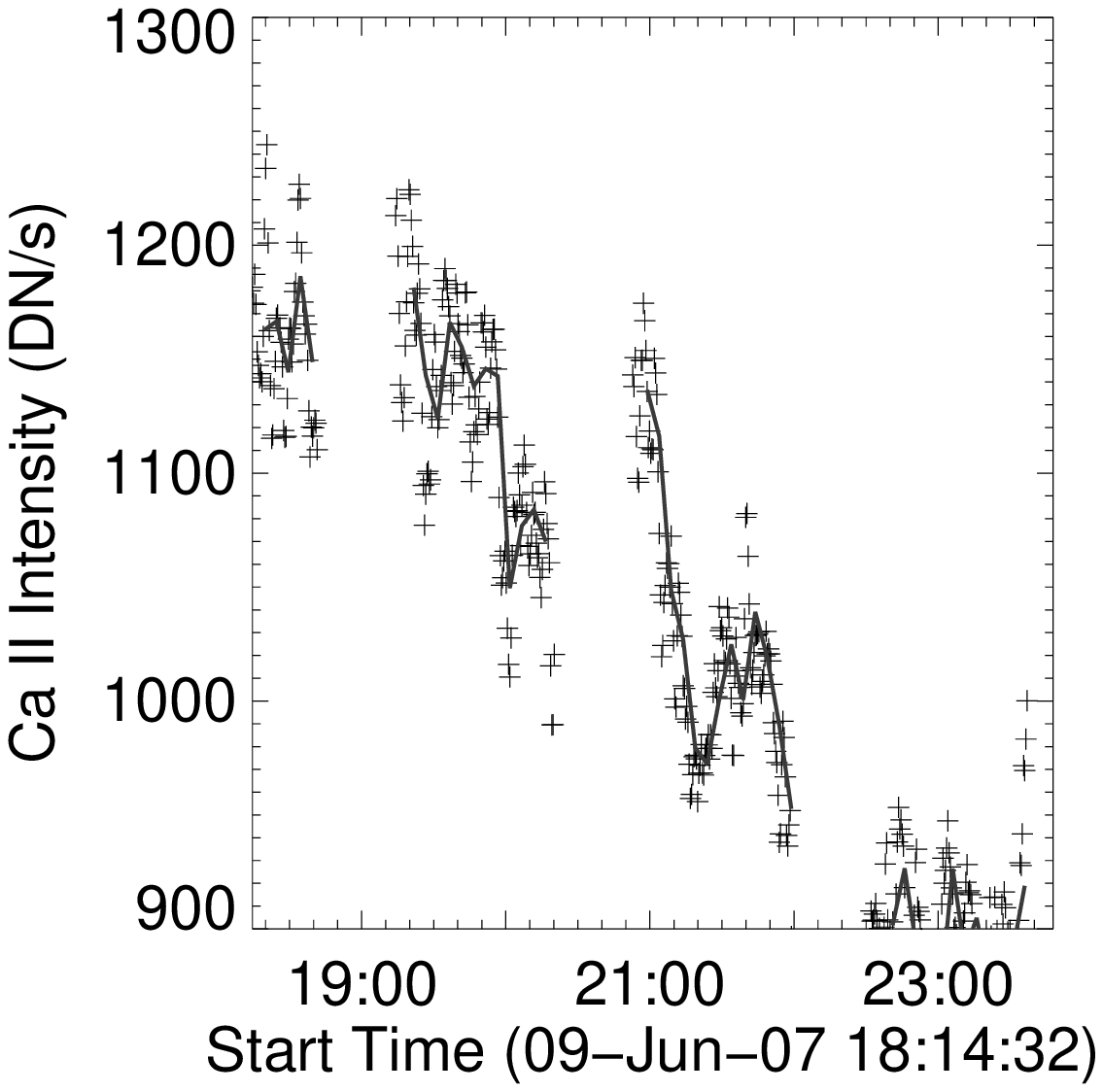}
\includegraphics[width=0.20\linewidth]{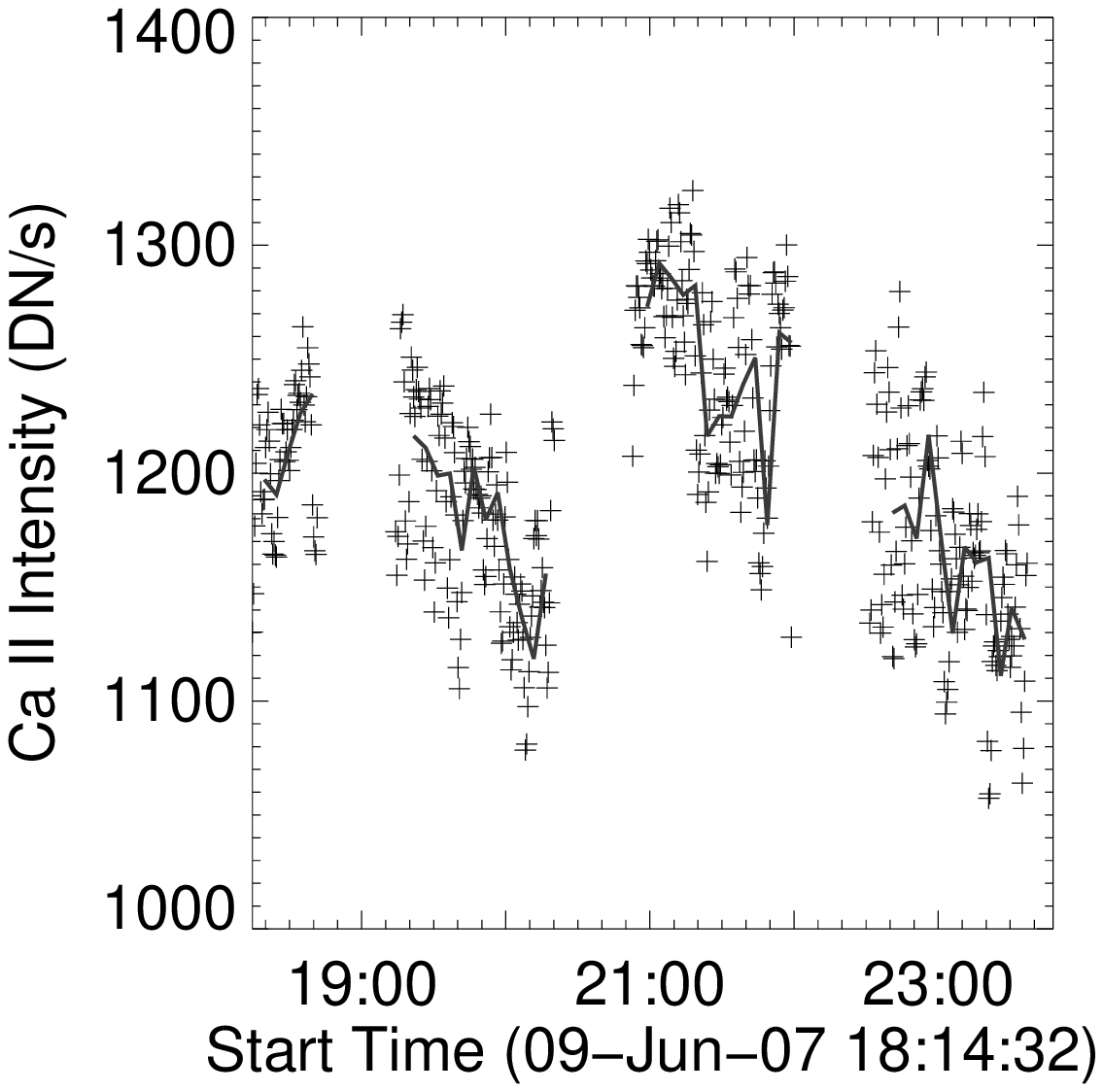}
\includegraphics[width=0.20\linewidth]{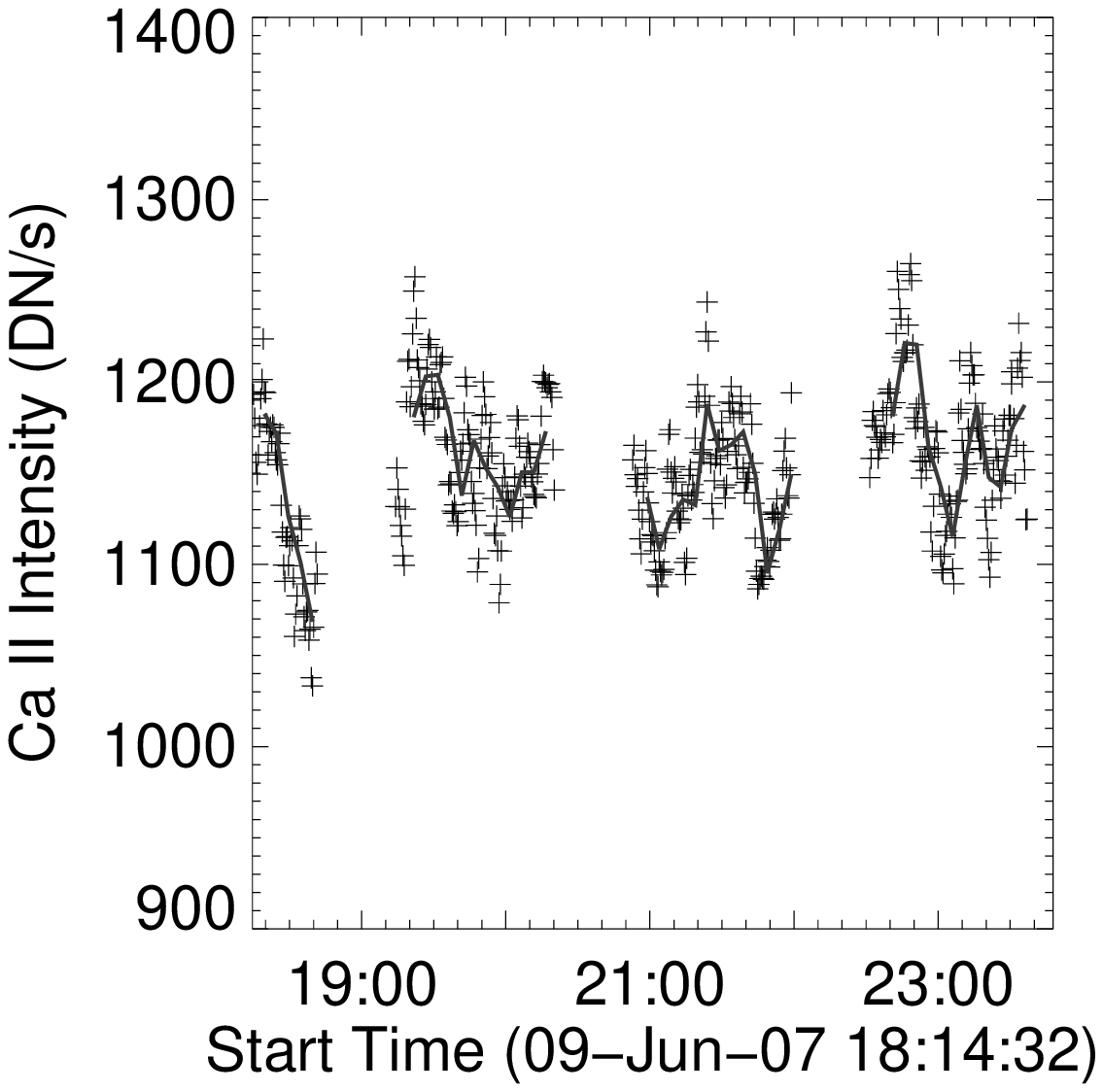}
\includegraphics[width=0.20\linewidth]{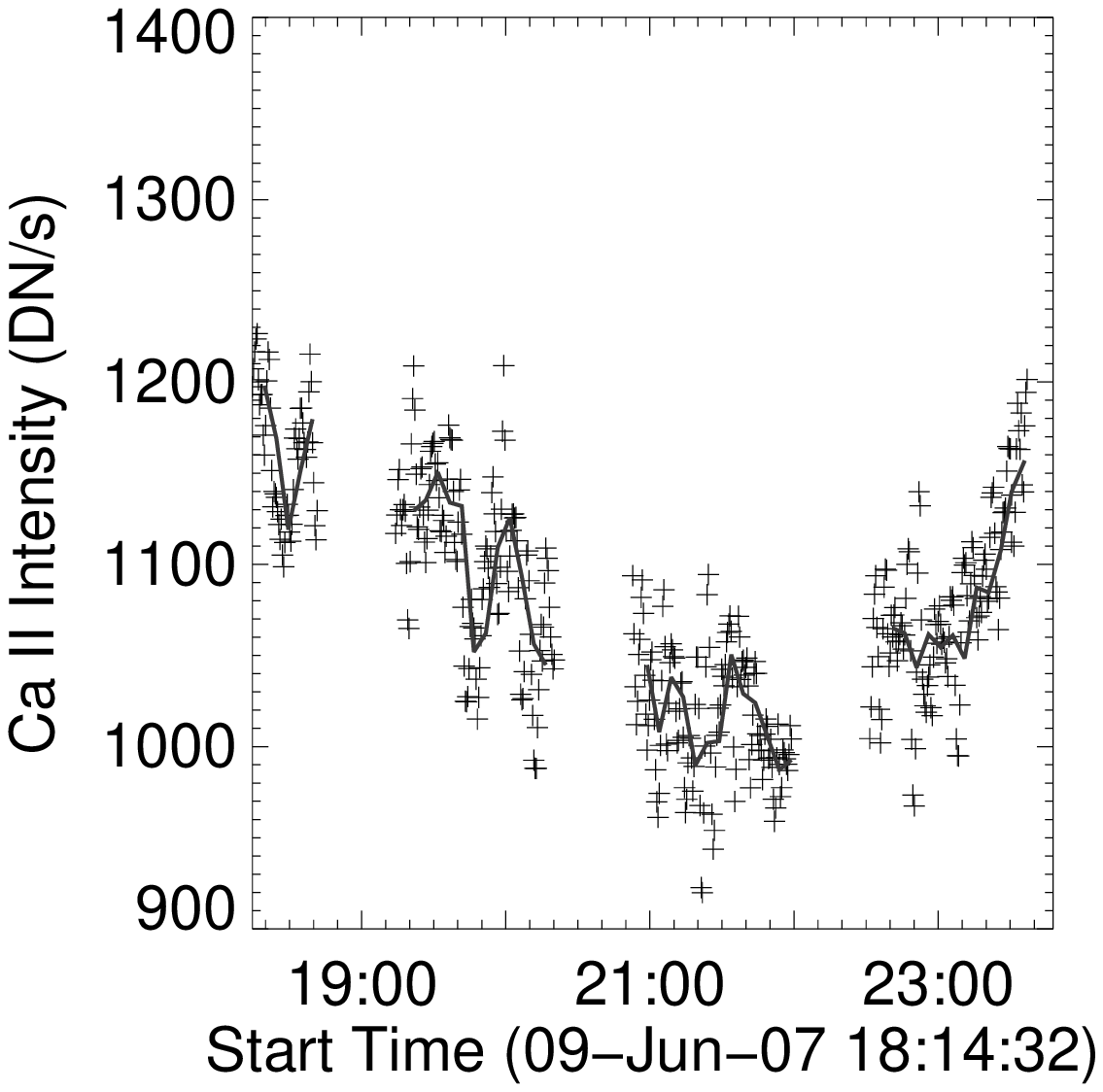}
\caption{Variation of moss \ion{Ca}{2} intensity (DN/s) as a function of time. Top left: standard deviation
of histograms of the pixel-to-pixel difference between successive images in the control box shown in Figure \ref{fig11}.
Other panels: evolution of the \ion{Ca}{2} inensity averaged over the small boxes in Figure \ref{fig11}. The crosses show the 
individual points, and the thick solid lines show detrended 5 mins averaged curves to de-emphasize the scatter.
\label{fig12}}
\end{figure*}

As with the magnetic flux analysis, we need to estimate the
statistical uncertainty in order to gauge whether variations in the \ion{Ca}{2} emission are significant. 
For this purpose we analysed a time-series of very high cadence (8s) \ion{Ca}{2} images of a plage region
obtained between 11:57 and 13:37UT on 2007, February 20, with an exposure time of 0.41s. 
The data were co-registered by cross-correlating
successive images and intensity difference images for the entire time-series were prepared. Histograms of the 
intensity differences over the full FOV (56$'' \times$ 28$''$) were then formed 
for each image in the time-series and fit with Gaussian functions to determine the standard deviation ($\sigma$).
The standard deviations are 
plotted as a function of time in the upper left panel of Fig \ref{fig12}. The average value is 9DN/s with only 
4\% variation in the standard deviations as a function of time. 
We adopt this average value as our estimate of the statistical noise.

Figure \ref{fig12} also shows the variation of intensity in the moss and inter-moss boxes. An animation (movie3.mpg)
shows the locations of the boxes, and their stability, during the time-series. Remarkably, the average intensities
in all the boxes (including the inter-moss box) are within 10\% of each other. Furthermore, the intensity variation
is less than 10\% in all the boxes for the duration of the observations ($>$ 5 hours). The largest variation can
be seen in the lower left hand box. As with the magnetic flux, this is again clearly a slow evolution from $\sim$ 1200DN/s at the start of
the sequence to $\sim$ 900DN/s at the end. Comparable magnitude variations of 100--200DN/s are again seen on time-scales
of tens of minutes, and these are significantly above the noise estimate. 

We computed the correlation coefficients between the \ion{Ca}{2} intensities and magnetic fluxes for each of the boxes.
Intriguingly, about half show a correlation coefficient $>$ 0.5, indicating a weak positive correlation and 
suggestive that the 100--200G changes in the magnetic flux do propogate energy and heating 
that leads to emission that is detectable above the noise level
in the chromosphere, though it is not universally seen. 
\citet{sakamoto_etal2008} have recently also reported the detection of fluctuations in 
{\it TRACE} 171\,\AA\, observations above the photon noise level that have durations that agree
well with their estimated loop cooling times, albeit for a different active region.  
Our analysis of EIS 195.119\,\AA\, observations of this
region (paper 1) showed only $\sim$ 15\% variations in intensity over many hours, but a detailed
comparison with the noise was not made. 
The lack of a consistent detectable signature in the chromosphere and a possible signature in the 
{\it TRACE} data could indicate
that the energy propogation is amplified in the corona, or
is in fact released there. 
Further work on this issue is warranted, including comparisons of intensity variations with numerical 
models \citep{antolin_etal2008}.

\subsection{Magnetic Field Vector}
\begin{figure*}
\centering
\includegraphics[width=0.88\linewidth,viewport= 50 40 432 216]{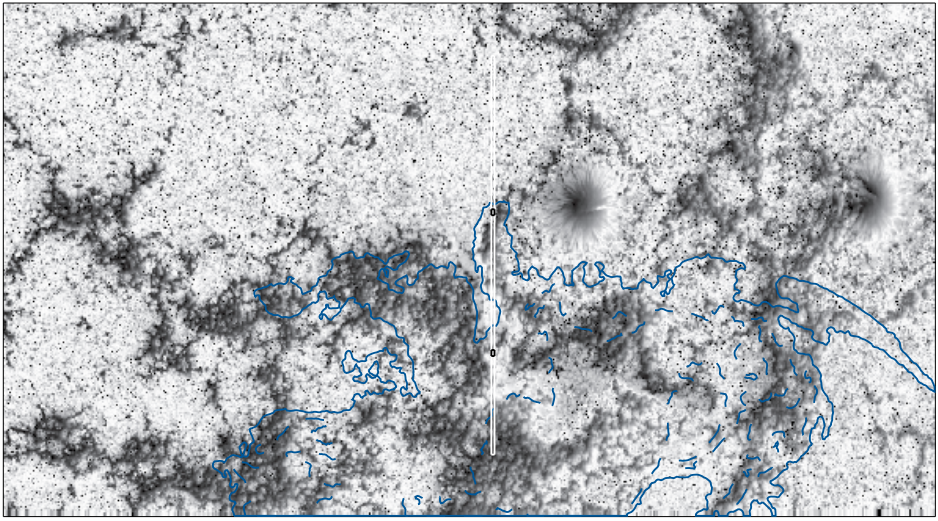}
\caption{Magnetic field line inclination at 14:20:05UT on June 09 with the FOV of the 
narrow SP scan overlaid in white. The small black boxes within the FOV show the areas chosen for 
analysis in Figure \ref{fig14}. The contours are the same as in Figure \ref{fig2}.
\label{fig13}}
\end{figure*}

Figure \ref{fig13} shows the same magnetic field inclination map represented by $f(\theta)$ 
as in Figure \ref{fig3} and defined in Equation \ref{eq1}.
The FOV of the narrow (7$'' \times$512$''$) SP slit scan from June 08 is overlaid. The
FOV has been rotated to the time of the large FOV scan (14:20 on June 09). The same 171\,\AA\,
contours as before are shown. It can be seen that the narrow scan crosses the inter-moss 
region South-North and reaches both positive and negative polarity moss regions. Both regions
also appear to be moderately inclined to the line-of-sight (black) whereas inter-moss region shows 
strong inclination (white). 
Two regions are selected  
in the moss (the small boxes within the FOV in Figure \ref{fig13}) to study the 
variation of the field line inclination and azimuth angle. The 
areas of the boxes are again 2$'' \times$2$''$. 

\begin{figure}[ht]
\centering
\includegraphics[width=0.98\linewidth]{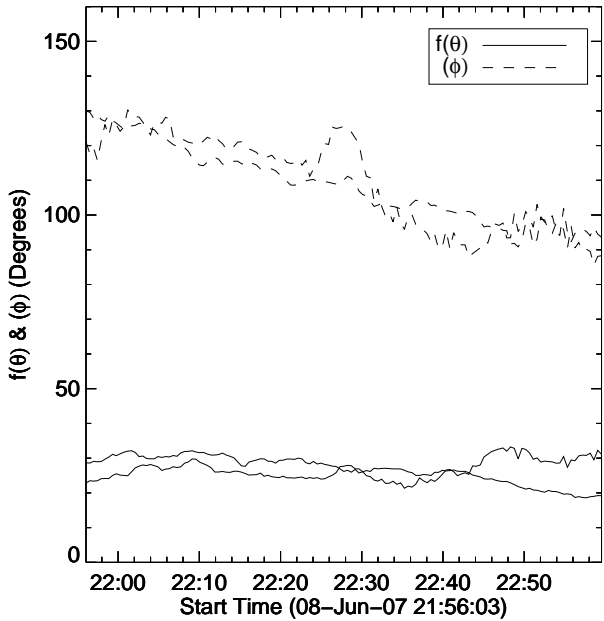}
\caption{Variation of magnetic field inclination and azimuth angle as a function of time for
the positive and negative polarity moss regions shown as boxes in Figure \ref{fig13}.
The inclination is represented by $f(\theta)$ (see text). 
Less than 15\% variation is seen in all quantities over a period of about 1 hr. 
\label{fig14}}
\end{figure}

Figure \ref{fig14} shows the results. Consistent with the measurements in \S \ref{characterB}
both moss regions show moderate inclination from the line of sight with an average value of 27$^o$. The
variation is less than 15\% over the time-series (21:56:03--22:59:31UT). The azimuth 
angle changes from $\sim$ 120$^o$ at the start of the time-series to $\sim$ 90$^o$ by the end, but as with
other quantities, this is a slow evolution, and the variation from the average ($\sim$ 109$^o$) is less than 15\% over the whole 
period.
Note that this time-series lasts for 3808s, which is again much longer than the theoretical 
cooling times for loops extrapolated from the moss regions calculated in \S \ref{model}.

One criticism of such simple analyses is that variability on
small scales is smoothed out by the averaging over large areas. 
It is possible that coronal emission is 
quasi-steady over many hours on size scales at the limit of current EUV/X-ray instrumentation, but that 
future instrumentation will find higher variability on smaller spatial scales. 
What tends to happen is that quantities appear relatively steady for many hours, within some threshold, when averaged over
large areas. These quantities also appear steady when averaged over smaller areas, but for
shorter periods of time, though the duration can be extended if the threshold is relaxed. 
For example, the most variable quantity in Figure \ref{fig14} is the inclination angle in the
positive polarity box. It varies by 
less than 15\% over the observations period. Individual pixels in this box,
however, have larger variabilities: $\sim$ 34\% on average. Relaxing 
our definitions slightly, however,
we are still able to say that $>$ 80\% of the individual pixels vary by less than 30\% over 
time-scales comparable to the cooling times for 85\% of the loops in our simulations (1920s). 

The key question of course is how these time-scales, spatial scales, and thresholds,
compare to the critical values for coronal heating. 
We have shown that the chromospheric \ion{Ca}{2}
emission and all the magnetic activity in
AR 10960 evolves slowly, and our quantitative analysis also indicates that the variability in magnitude is only at a low
level on time-scales longer
than theoretical loop cooling times in small 2$'' \times$2$''$ boxes.
The movies presented in this paper clearly suggest, however, 
that there is a lot of small-scale dynamic activity.
The time-scale for loop cooling is clearly resolved in these observations, but the important spatial
scales may not be. Next we investigate the moss filling factor in this region using EIS observations in order to 
at least
indirectly infer the size scale of a fundamental loop envelope. 

\section{EIS Filling Factor Measurements}
\label{ff}
Previous studies have shown that intensities in the moss scale linearly with the loop base
pressure and independently of the loop length \citep{martens_etal2000,vourlidas_etal2001}. 
As pointed out by \citet{warren_etal2008b}, this means that any 
physical model that yields the same base loop pressure will predict
the same moss intensity regardless of the loop length. \citet{warren_etal2008b} calculated 
full solutions to the hydrodynamic 
equations, assuming steady heating, for a grid of loop lengths from 10--100Mm.
This grid sufficiently covers most of the distribution of loop lengths we computed for the moss 
in \S \ref{model}. For each loop solution, the density and temperature around the loop is known
and the intensity at each position for any EIS line can be calculated. By integrating the intensity
over the lower 5Mm of the loop the footpoint intensity can be computed. The intensities for the 
grid of loops scale linearly with pressure and can be fitted with a function of the form $I_\lambda 
= a P_0^b$, where $I$ is the intensity and $P_0$ is the base pressure. \citet{warren_etal2008b}
state that the linear relationship between intensity and pressure breaks down at low pressures,
so their fits are restricted to values above $\log$ P$_e$ = 16 cm$^{-3}$ K.

The intensity ratios of 
selected lines can also be directly related to the simulated base pressure.
By determining the moss pressure from observed intensity ratios, a simulated 
intensity for an individual line can be calculated from the fit and compared to the observed intensity
for that line. The filling factor
is then introduced (if needed) to bring the simulated and observed intensities into agreement.
We followed this procedure here for AR 10960.

%%%%%
%%%%
Following the analysis in paper 2 we use a density threshold to identify the moss pixels.
Several density diagnostics are available in the EIS wavelength bands and a detailed discussion
of comparisons between them has been presented in \citet{young_etal2009}. In paper 2 we discussed these
comparisons 
and decided to use the \ion{Fe}{13} 202.044/203.826\,\AA\, ratio
because of the consistency between densities derived from this ratio and densities derived from 
\ion{Si}{10} ratios in small active regions, bright points, and the quiet Sun i.e. regions where the
density is lower than in the moss and the \ion{Si}{10} lines are sensitive. For
comparison with that work, we use this ratio again here. The analysis is applied to the EIS slit 
raster scan taken at 10:58:10UT on June 09, and 
discussed in \S \ref{obs}. 
The \ion{Fe}{13} 202.044\,\AA\, line was fitted at every 
%%%%
pixel with a single Gaussian.
The \ion{Fe}{13} 203.826\,\AA\, line is a self-blend of two lines and is 
further blended at 203.734\,\AA\, with an \ion{Fe}{12}
line \citep{brown_etal2008}. We performed a double Gaussian fit to this feature at every pixel to 
remove the \ion{Fe}{12} blend. To do this, we fixed the separation of the two components at 0.1\,\AA\,
and forced them to have the same width. 

\begin{figure}[ht]
\centering
\includegraphics[width=0.48\linewidth, viewport= 100 20 300 360]{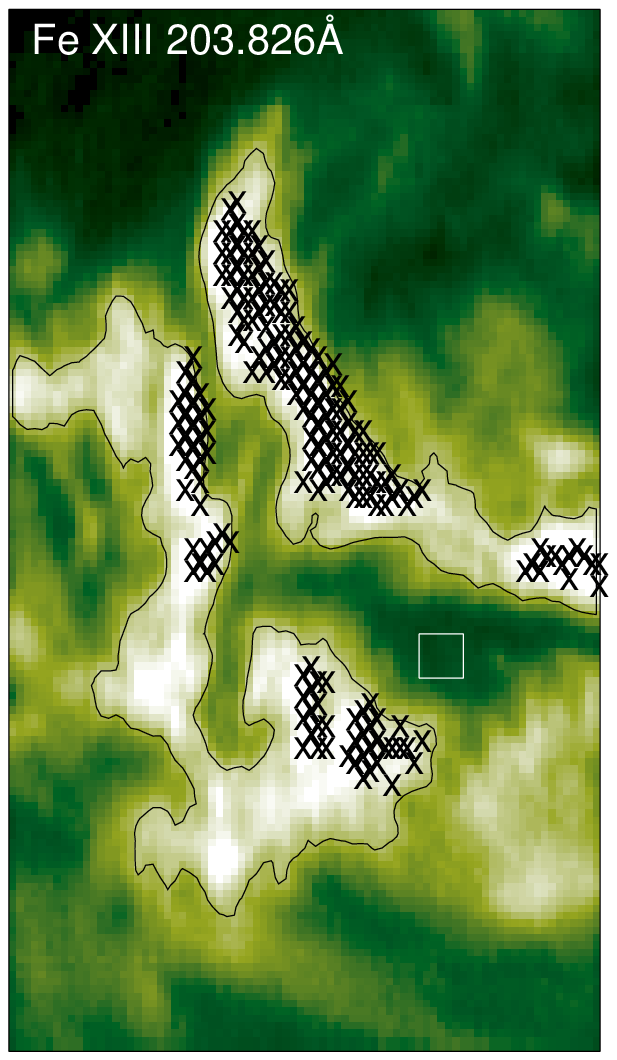}
\includegraphics[width=0.48\linewidth, viewport= 100 20 300 360]{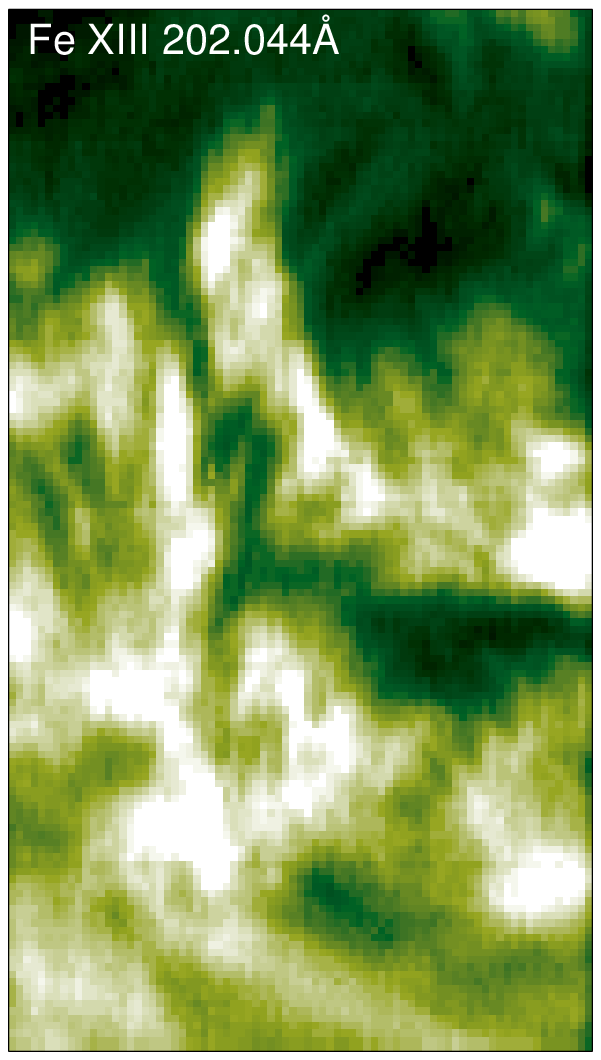}
\caption{EIS images of the core of AR 10960 formed from fits to the density sensitive \ion{Fe}{13}
lines. The moss regions are highlighted by a contour at 40\% maximum intensity. The small white
box shows the inter-moss region used for background subtraction, and the crosses show the moss
pixels with $\log$ N$_e > $ 9.6 cm$^{-3}$ used for determining the filling factor.
\label{fig15}}
\end{figure}

\begin{figure}[ht]
\centering
\includegraphics[width=0.98\linewidth]{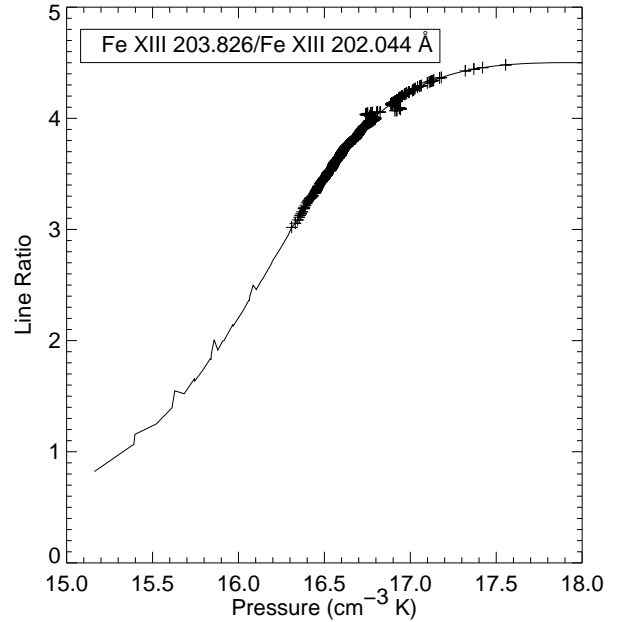}
\caption{Theoretical line ratio versus base pressure curve for the grid of loop models discussed
in the text. The values for the moss pixels highlighted in Figure \ref{fig15} are overplotted as
crosses.
\label{fig16}}
\end{figure}

Images of the active region core formed from the line fits are shown
in Figure \ref{fig15}. The full raster spanned several passes through {\it Hinode} night,
so the images only show a central area of 80$'' \times$140$''$. 
Electron densities were calculated for these
regions using the measured line ratios and the CHIANTI database version 6.0 \citep{dere_etal1997,dere_etal2009}.
A contour set at 40\% of the maximum intensity is drawn on the 
\ion{Fe}{13} 203.826\,\AA\, image of Figure \ref{fig15} to highlight the moss regions. 
Pixels within 
these contours were selected
and identified as moss if the calculated electron density exceeded $\log$ N$_e$ = 9.6 cm$^{-3}$.
These are shown as crosses on the \ion{Fe}{13} 203.826\,\AA\, image.
For each of the identified pixels, the observed intensity ratio was computed after subtracting
contaminant background emission. This background intensity was measured in the small box
in the inter-moss region on the \ion{Fe}{13} 203.826\,\AA\, image. The new ratios were then
used to derive the base pressure by interpolation from the simulated grid of solutions.
Figure \ref{fig16} shows the values for the moss pixels overplotted on the theoretical
line ratio vs base pressure curve. Although several moss pixels approach the high pressure
limit of the ratio, it can be seen that the vast majority of them fall in the sensitive range
of the curve. The minimum base pressure in the moss is $\log$ P$_e$ = 16.3 cm$^{-3}$ K, which
is higher than the lower limit of the power law fits of \citet{warren_etal2008b}.
With the base pressure established, the \ion{Fe}{13} 203.826\,\AA\, intensity was simulated
using the fit for \ion{Fe}{13} 203.826\,\AA. The coefficients
of the fit in this case are $a$ = -11.94 and $b$ = 0.99.

\begin{figure}[ht]
\centering
\includegraphics[width=0.98\linewidth]{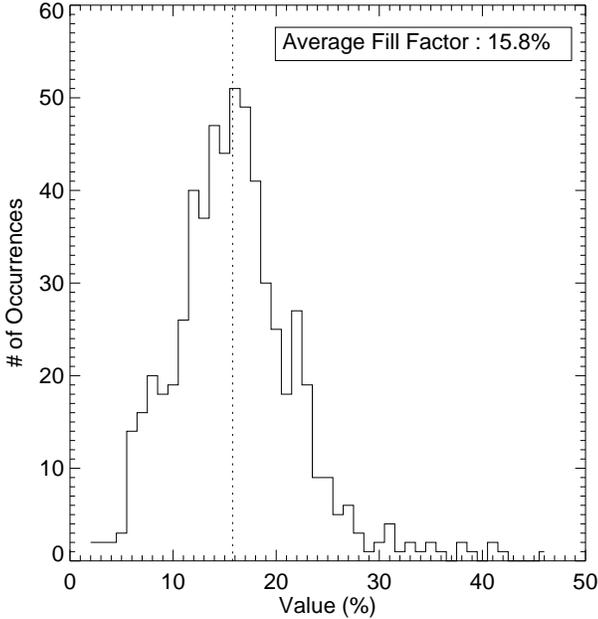}
\caption{Distribution of filling factors computed for the moss pixels.
\label{fig17}}
\end{figure}

As expected, the intensities thus simulated are much higher than observed, so a filling factor
needs to be introduced to bring them into agreement. The distribution of filling factors for
the moss pixels is shown in Figure \ref{fig17}. The majority of values fall in the 10--20\% range
with the median value being $\sim$16\%. This is in agreement with previous measurements of moss
filling factors \citep{fletcher&depontieu_1999,warren_etal2008b}.

These results suggest that the fundamental size scale of structures in the moss are about 10--20\% of
the EIS spatial resolution. This is consistent with similar measurements of coronal loops
with EIS based purely on observational factors \citep{warren_etal2008a}, and also sets an upper limit
on the cross-field scale of coronal heating, which will be much smaller if loops are composed of
multiple threads. Note that the filling
factor is an area filling factor. EIS 
has 1$''$ spatial pixels, so taken at face value a next generation instrument needs 0.3--0.4$''$
spatial pixels to resolve the median moss filling factor. Interestingly, this is larger than
the size of the SOT FG and SP pixel scales. Although EIS (and XRT) cannot directly resolve these scales,
SOT can resolve these scales below the moss. 

\section{Magnetic flux evolution and chromospheric emission in individual pixels}
\label{individual}
Having inferred the apparent cross-field size scale of fundamental structures in the moss and established that the FG can
resolve this scale in individual pixels, we revisit our analysis of the variability in magnetic
flux and chromospheric emission in the moss region (\S \ref{variability}). 
Previously we showed that the average magnetic field
in most of the boxes 
varied by $\sim$ 15--30\% while the chromospheric \ion{Ca}{2} 3896\,\AA\, intensity varied by $\sim$10\%.
Note that these changes include changes due to motions in and out of the field of view as well as 
variations in magnitude. These motions may have a more pronounced effect on the results for individual pixels
because the features can traverse the pixel size scale faster. It is worth pointing out that features
moving in and out of a pixel can have the effect of canceling each other to some degree so that variability
may be reduced. This would
also be true if the fundamental structures are composed of multiple threads.

%%%%%
\begin{figure}[ht]
\centering
\includegraphics[width=0.98\linewidth]{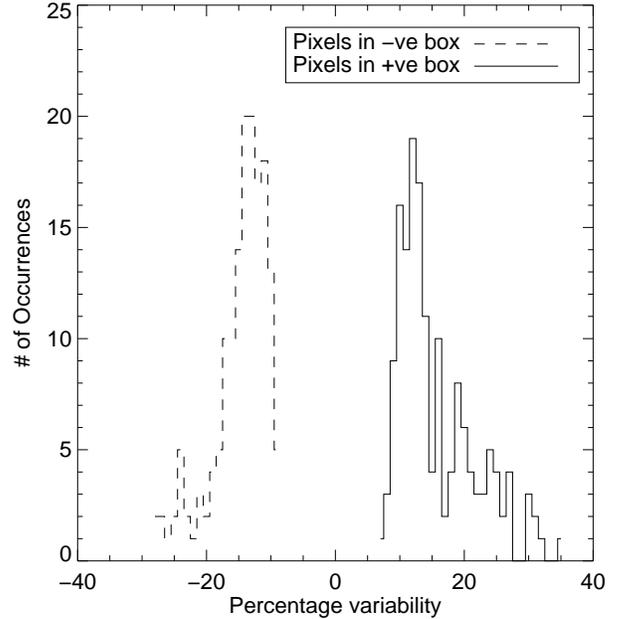}
\caption{Distributions of magnetic field variability ($\sigma$/$\bar{B}$) expressed as percentages
for the individual pixels
within a positive and negative polarity moss box from Figure \ref{fig9}.
\label{fig18}}
\end{figure}

Figure \ref{fig18} shows the distribution of percentage variabilities 
(ratio of standard deviation $\sigma$ to average magnetic
field $\bar{B}$) for all the pixels in two of the boxed regions. Each box contains 156 pixels.
The values are calculated for a time-period of 1920s. Recall that 85\% of the loops in
our simulation cool on shorter time-scales than this. In both cases $>$ 75\% of
the pixels show variabilities less than 20\% and the median values for both distributions
are $\sim$ 13\%. 
The variability
is clearly greater in individual pixels over the full observations period, 
but this is also partly attributable to the lifetime of
individual features. For the record, more than 2/3 of the pixels in the boxes show less than 45\% variability
for the whole $>$ 5 hr period.

\begin{figure}[ht]
\centering
\includegraphics[width=0.98\linewidth]{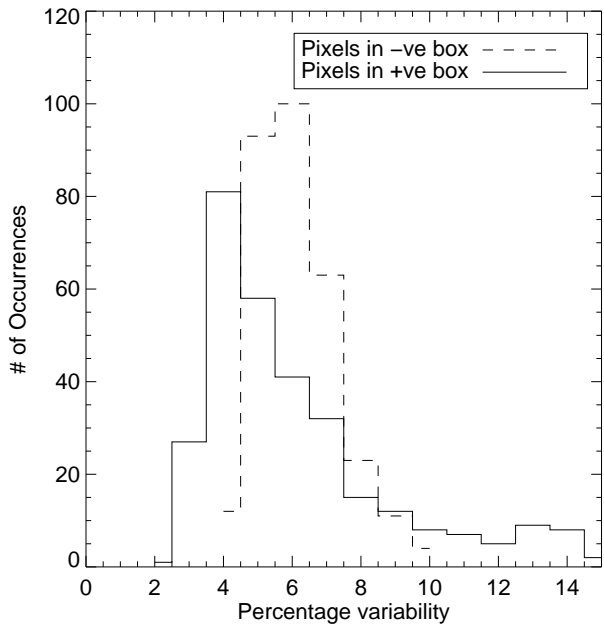}
\caption{Distributions of \ion{Ca}{2} intensity variability ($\sigma$/$\bar{I}$) expressed as percentages
for the individual pixels
within the same two moss boxes shown in Figure \ref{fig18}.
\label{fig19}}
\end{figure}

Figure \ref{fig19} shows the distribution of percentage variabilities of the \ion{Ca}{2} 3896\,\AA\, intensities
for all the pixels in the same two boxed regions, computed for the same time-period. 
In this case, each box contains 306 pixels. In both cases $\sim$ 85\% of the pixels show
variabilities of less than 10\% and the median values for both distributions
are $\sim$ 6\%. Again, the variability is larger over the full observations period but, at least for the selected
boxes, all the pixels show less than 15\% variability for the whole $>$ 5 hr period.

\begin{figure}[ht]
\centering
\includegraphics[width=0.98\linewidth]{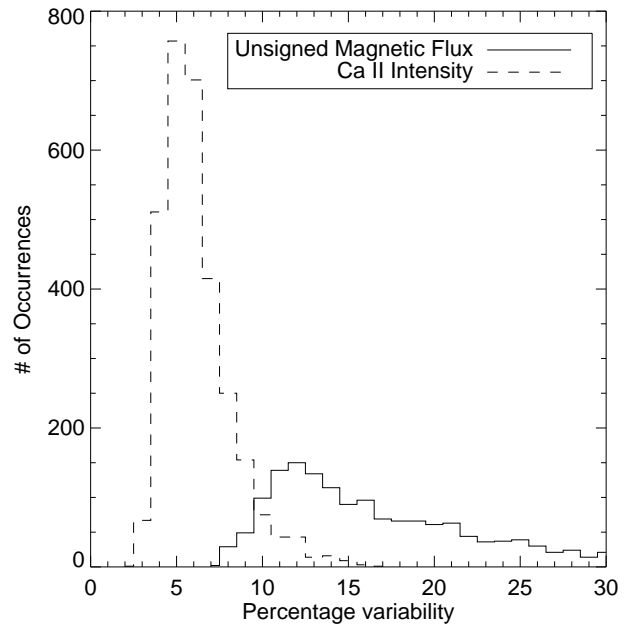}
\caption{Distributions of \ion{Ca}{2} intensity variability ($\sigma$/$\bar{I}$) and magnetic field variability ($\sigma$/$\bar{B}$) expressed as percentages
for all the pixels in all the moss boxes shown in 
Figure \ref{fig9}.
\label{fig20}}
\end{figure}

To check whether the two examples are representative, we show in Figure \ref{fig20} the distribution
of percentage variabilities of the \ion{Ca}{2} 3896\,\AA\, intensities and unsigned magnetic
flux for all individual pixels within all the moss boxes. The \ion{Ca}{2}
distribution has a median around 6\% with 99\% of the pixels showing variabilities of less than 15\%.
The median of the magnetic flux distribution is around 15\% with more than 2/3 of the pixels 
showing variabilities of less than 20\%. These values were again computed for a time-period of
1920s. As discussed in \S \ref{variability}, the magnetic flux in the moss shows greater variation,
but this does not seem to register strongly in the chromosphere.

%%%%%
%%%%%
%%%%%

\section{Discussion}
\label{discuss}
We have described the characteristics and evolution of the longitudinal and vector magnetic field and 
chromospheric \ion{Ca}{2} emission in the core
of an active region observed by {\it Hinode} and {\it TRACE}. Consistent with previous studies, we
found that the moss in this region is unipolar, 
the field lines are only moderately inclined, and the spatial distribution of magnetic flux in the core
evolves slowly. We put the last statement on a quantitative basis by comparing the evolution time-scales
with the theoretical cooling times computed from hydrodynamic simulations of coronal loops extrapolated from
SP magnetograms of the moss regions. These extrapolations also showed that the magnetic connectivity in the
moss is different than in the quiet Sun because most of the magnetic field extends to significant coronal
heights (there is nothing to connect to locally in the unipolar regions). We also
showed that the field line inclination and horizontal component are coherent in the moss, and most of the
shearing of the field occurs in the inter-moss region, or at the edges of the moss, where the magnetic
flux variability was shown to be greatest. The 
magnetic vector, flux, and \ion{Ca}{2} intensity in 2$'' \times$2$''$ boxes in the moss also do not show significant variability
on time-scales
longer than loop cooling time-scales when averaged over these areas. Though flux changes on the order of 100-200G 
and \ion{Ca}{2} intensity changes on the order of 100-200DN/s
are observed on 20--30 mins time-scales, we find only weak evidence that these flux changes consistently register in
the chromosphere.
We also determined the filling factor in the moss from EIS observations: $\sim$16\%. This is consistent with 
previous studies, and suggestive that the cross-field scale of the fundamental structures in the moss are larger than the   
size of an individual SOT FG pixel (0.16$''$).
The mean variabilities of the magnetic flux and chromospheric emission
in individual FG pixels was also found to be $\sim$15\% and $\sim$6\%, respectively, on time-scales longer
than the computed loop cooling times. 

EUV and X-ray observations of the moss in the core of this region have already suggested that the heating is 
steady or effectively steady i.e. heating events occur with a rapid repetition rate.
The results presented here show that the magnetic field and chromospheric emission also evolve slowly,
and remain relatively steady even at the highest spatial resolution we have
ever observed. They could therefore be 
interpreted as further evidence supporting the quasi-steady picture. 
The short time-scale 100-200G variations we see could be interpreted as evidence of low frequency
impulsive heating. We find, however, that only a small fraction of our simulated loops (10\%) would
be expected to cool on these time-scales.

If the heating events do occur frequently and are reconnection related,
one would also expect to see changes in the magnetic field on a comparable time
scale.
Further studies of the relationship between the detailed small scale changes in the magnetic flux that we see
and variations in the EUV/X-ray intensities
are needed to address this issue. At present it is unclear theoretically how important changes of 10--20\% are
in terms of energy input. The magnetic field inclination, for example, is indicative of the horizontal
stress component of the field and could be directly realted to the heating. 
Small variations of this angle do not necessarily imply that the heating variations are also small.
One nuance is that the measurements of changes in the magnetic field/flux are a 
combination of changes due to translational motions as well as variations in magnitude, so direct
conversion into energy is not unambiguous. Nevertheless,      
the quiet Sun chromospheric and coronal heating requirements are 
$\sim$4$\times$10$^{6}$ erg cm$^{-2}$ s$^{-1}$ and
$\sim$3$\times$10$^{5}$ erg cm$^{-2}$ s$^{-1}$, respectively
\citep{withbroe&noyes_1977,aschwanden_2004}, suggesting that 10\% changes in chromospheric
output could be enough to sustain the quiet corona. The situation is more complex, however, in active
region loops. \citet{kano&tsuneta_1996} and \citet{katsukawa&tsuneta_2005}
quoted estimates of 
$\sim$10$^{6}$ erg cm$^{-2}$ s$^{-1}$ for `warm' loops and 
$\sim$10$^{7}$ erg cm$^{-2}$ s$^{-1}$ for hot loops. This compares to 
$\sim$2$\times$10$^{7}$ erg cm$^{-2}$ s$^{-1}$ for the chromosphere in active regions
\citep{withbroe&noyes_1977,aschwanden_2004}. Assuming that the strong basal levels of
magnetic flux and \ion{Ca}{2} emission in AR 10960 match this heating requirement, these 
numbers suggest that the 10--20\% changes we observe 
on `warm' EUV loop
cooling time-scales of 20--30 mins could be enough to heat these loops impulsively at low frequency. 
These changes, and the 10--20\% variations that occur over time-scales that
are longer than a loop cooling time, would not be enough, however, to heat
the hot loops.

%%%%
%%%%
%%%%
\citet{katsukawa&tsuneta_2005} estimated the energy flux into coronal loops as a result of the
dissipation of magnetic energy built up in coronal current sheets due to the braiding of the field
by photospheric motions. They found that for a moss magnetic field strength measurement of 1.2kG, the 
heating requirement of hot loops (10$^{7}$ erg cm$^{-2}$ s$^{-1}$) could be met. In their estimates,
however, the energy flux is related to the magnetic flux as 
$F \propto B^2$, so that once again a 10--20\% change in magnetic flux may meet the heating 
requirements of the `warm' loops, but does not appear to supply enough energy to heat the hot loops.
A definitive statement on this issue awaits more detailed
theoretical modeling. These results suggest, however, that only
the continuous input from the strong basal levels of magnetic flux and chromospheric emission,
not the variations,
appear to be enough to power the hot core loops rooted in the moss.

The results presented here, together with those in Paper 1 and Paper 2
and \citet{antiochos_etal2003},
show a lack of variability in many types of diagnostic signatures: 
magnetic flux observations by SOT, 
coronal velocity measurements by EIS, EUV and X-ray intensities from {\it TRACE} and XRT, etc., and constitute
a compelling body of evidence.
Future instrumentation, however, may reveal all of these quasi-steady 
properties to be a distraction. 
It is possible, for example, that the magnetic field could appear steady in the lower atmosphere while it is braided
in the corona and energy is released there. Heating by impulsive events could occur at coronal heights 
with no visible signature  
in the magnetic field at lower heights. The lack of variability in the EUV and X-ray data could indicate that
the impulsive events occur on sub-resolution scales.
Low frequency impulsive heating on unresolved scales could give the impression of high frequency
(effectively steady) heating.
These suggestions, of course, are difficult to test, since we do not know the spatial
scale of the heating. \citet{priest_etal2002} suggest that the size scale of the fundamental kG flux tubes in the
photosphere have diameters of 100km, and based on these arguments \citet{klimchuk_2006} suggests that a
comparable spatial resolution may be able to resolve the strands making up the corresponding loops,
though the current sheet interface may be smaller.
We have shown here that the magnetic flux and chromospheric emission are quasi-steady on spatial scales approaching 100km.
This is 4000 times smaller than studied by \citet{antiochos_etal2003} and should rule out models that predict
low frequency impulsive heating on larger scales. 

\acknowledgments
We would like to thank Jim Klimchuk, Saku Tsuneta, Tom Berger, and Alfred de Wijn for very helpful discussions.
DHB and HPW acknowledge funding support
from the NASA {\it Hinode} program. The EIT images are courtesy of the {\it SOHO} EIT Consortium. {\it SOHO} is a mission of international cooperation
between ESA and NASA. The {\it TRACE} images are courtesy of the {\it TRACE} Consortium.
Hinode SOT/SP Inversions were conducted at NCAR under the framework of the Community Spectro-polarimetric Analysis Center (CSAC; \url{http://www.csac.hao.ucar.edu/}). 
CHIANTI is a collaborative project involving researchers at NRL (USA), RAL (UK), and the Universities of: Cambridge (UK), George Mason (USA), and Florence (Italy).
{\it Hinode} is a Japanese mission developed and launched by ISAS/JAXA, collaborating with NAOJ as a domestic partner, NASA and STFC (UK) as international partners. Scientific operation of the {\it Hinode} mission is conducted by the {\it Hinode} science team organized at ISAS/JAXA. This team mainly consists of scientists from institutes in the partner countries. Support for the post-launch operation is provided by JAXA and NAOJ (Japan), STFC (U.K.), NASA, ESA, and NSC (Norway).

{\it Facilities:} \facility{Hinode (EIS,SOT), TRACE, SOHO (EIT,MDI) }     

\bibliography{/data2/latex/dhb_bib/solar}
\end{document}